\documentstyle[12pt]{article}

\renewcommand{\theequation}{\thesection.\arabic{equation}}
\def\bea{\begin{eqnarray}}
\def\eea{\end{eqnarray}}
\def\bc{\overline{c}}

\begin{document}
\begin{flushright}
gr-qc/0102104
\end{flushright}

\begin{center}{\Large {\sf Gravitating Self-dual Chern-Simons Solitons
}}\\[10mm]
Bok Keun Chung and Jin-Mo Chung\\
{\it Asia Pacific Center for Theoretical Physics, Seoul 130-012, Korea}\\
{\it and}\\
{\it Research Institute for Basic Sciences and Department of Physics,}\\
{\it  Kyung Hee University, Seoul 130-701, Korea}\\
{E-mail: bkchung$@$khu.ac.kr, jmchung$@$khu.ac.kr}\\[2mm]
Seongtag Kim\\
{\it Department of Mathematics and Institute of Basic
Science, Sungkyunkwan University,}\\
{\it Suwon 440-746, Korea}\\
{E-mail: stkim$@$skku.ac.kr}\\ [2mm]
Yoonbai Kim\\
{\it BK21 Physics Research Division and Institute of Basic Science,
Sungkyunkwan University,}\\
{\it Suwon 440-746, Korea}\\
{E-mail: yoonbai$@$skku.ac.kr}
\end{center}

\vspace{5mm}

\begin{center}
{\sf Abstract}\\[2mm]
\end{center}
\begin{quote}
\hspace{7mm}
Self-dual solitons of Chern-Simons Higgs theory are examined in curved
spacetime. We derive duality transformation of the Einstein Chern-Simons Higgs
theory within path integral formalism and study various aspects of dual
formulation including derivation of Bogomolnyi type bound.
We find all possible rotationally-symmetric soliton configurations carrying
magnetic flux and angular momentum when underlying spatial manifolds of these
objects comprise a cone, a cylinder, and a two sphere.
\end{quote}

\newpage

\setcounter{equation}{0}
\section{Introduction}
Since the Chern-Simons gauge field made appearance in physics
literature as a mass term in three dimensional gauge
theory~\cite{JT}, it has attracted the attention in various ways,
e.g., anyon and fractional statistics~\cite{ASWZ}, a relation to
knot theory~\cite{Wit}, and conformal field theories in two
dimensions~\cite{BN}. Chern-Simons gauge theories coupled to
scalar matter field also support the soliton solutions carrying
fractional angular momentum~\cite{WZ,PK}. An example which
provides static multi-soliton solutions is relativistic
Chern-Simons Higgs model of a specific $\phi^{6}$ scalar
potential, and it admits both topological and nontopological
solitons~\cite{HKP,JLW,Dun}. For the vortices in Abelian Higgs
model with critical coupling, it was proved in Ref.~\cite{CG}
that the Bogomolnyi type bound is saturated after the inclusion
of Einstein gravity. The progress to this direction has also been
achieved in Chern-Simons Higgs model coupled to background
gravity~\cite{Sch} and Einstein gravity~\cite{Val}. The solitons
of Abelian Higgs model are distinguished from those of
Chern-Simons gauge theories by the fact that whether they are
spinning objects or not. This angular momentum affects the
derivation of Bogomolnyi bound in curved space: the Bogomolnyi
limit of Nielsen-Olesen vortices is obtained under the static
metric and the same $\phi^{4}$ scalar potential as  in flat
space case, however that of the self-dual Chern-Simons solitons is
saturated under the general stationary metric and thereby
introduce a $\phi^{8}$ scalar potential of negative coefficient
in addition to the $\phi^{6}$ potential of flat
spacetime~\cite{Val}. Furthermore, the scalar potential in the
Bogomolnyi limit of the Einstein Chern-Simons Higgs model
includes, at least, one parameter, even if we use the condition
that the obtained first-order equations reproduce original
second-order Euler-Lagrange equations and their solutions carry
finite magnetic flux (or equivalently charge) and angular
momentum.

In this paper, for the static configurations of Chern-Simons Higgs model, we
shall derive conventional Bogomolnyi type bound of the total energy defined
by Euler number in terms of magnetic flux, the topological charge of the
system. Examining the obtained Bogomolnyi equations by use of both analytic
and numerical methods, we shall show that there exist only two types of smooth
solutions despite the complicated $\phi^{8}$ potential unbounded below.
They are topological vortices and nontopological solitons which coincide
exactly with the solitonic spectra in flat spacetime.
First, when boundary value of the scalar field has a Higgs vacuum, self-dual
vortex solutions are supported and the underlying manifolds of these solutions
are asymptotic cone and cylinder. Second, when boundary value of the scalar field
has a symmetric local minimum, self-dual nontopological solitons are produced
and the corresponding spatial manifold constitutes two sphere, asymptotic
cylinder in addition to asymptotic cone. It is an intriguing point that the
scalar potential vanishes at the boundary value of both solitons in open
spatial manifold, which implies zero cosmological constant at asymptotic
open space. Examining asymptotic behaviors of solutions for multi-solitons,
we conclude that there exists an upperbound for the vorticity to form a vortex
and the obtained solutions are decaying fast as the radial distance increases.

A way to envisage the role of topological excitations in the path integral
formalism of Abelian gauge theories is to reformulate the given theory
through the dual formulation. It has been studied for the lattice version
of the Abelian Higgs model~\cite{ES}, and for the continuum version of the
Chern-Simons Higgs model~\cite{KL}. In this paper, we shall rewrite the
Chern-Simons Higgs model coupled to gravity in terms of dual gauge field.
In the dual transformed theory, we obtain the explicit forms of 
the nonpolynomial interaction
between the dual gauge and Higgs fields, and the topological interaction
between the dual gauge field and the topological sector of the scalar phase.
Furthermore, we demonstrate a
role of those interactions in order to produce nonperturbative excitations
and their
mutual interactions. For gravitational field, the path integral measure
of the duality transformed theory possesses a Jacobian relative to that of
original theory.

In the next section, we will introduce the Chern-Simons Higgs
model in curved spacetime and rederive the Bogomolnyi type bound.
In section 3, we rewrite our model through the duality
transformation and discuss physics of the topological excitations
comparing with the original theory. In section 4, we examine
the Bogomolnyi equations and obtain all possible regular
rotationally-symmetric solutions. We show that the decaying
property of solutions of the Bogomolnyi equations and
nonexistence of solutions with prescribed vortices in section 5.
Conclusions with some comments about our results are presented in
section 6.

\setcounter{equation}{0}
\section{Self-dual Chern-Simons Solitons in Curved Space}
In this section, we recapitulate the derivation of the Bogomolnyi
bound for Chern-Simons solitons in curved space. In order to obtain such
Bogomolnyi-type bound, we consider the energy defined by the spatial
integration of two dimensional scalar curvature instead of the matter action
for static configurations as has been done in Ref.~\cite{Val}.

The action for an Abelian Chern-Simons gauge field theory with Higgs mechanism
is written in curved spacetime
\begin{eqnarray} \label{action}
S&=&\int d^3\! x\,\,\!\!{\sqrt g } \left[\, -\,\frac{1}{16\pi G}R
+\frac{\kappa}{2}\frac{\epsilon^{\mu\nu\rho}}{{\sqrt g
}}A_{\mu}{\partial_\nu}{A_\rho}+\frac{1}{2}
g^{\mu\nu} \overline{D_\mu \phi} D_\nu \phi - V(|\phi|) \,\right],
\end{eqnarray}
where $\phi=|\phi| e^{i\Omega}$, gauge covariant derivative is
$D_\mu =\partial_\mu-ieA_\mu$, and scalar potential $V(|\phi|)$ will be fixed
to give a suitable Bogomolnyi type bound. Dynamics of
gravitation is governed by 2+1 dimensional Einstein gravity and the metric
takes the stationary form, which is compatible with static spinning objects
\begin{equation} \label{metR}
ds^2 = N^2 (dt + K_i dx^i )^2 - \gamma_{ij} dx^i dx^j .
\end{equation}
In the above equation the components of the metric, $N, K_i , \;{\rm and}\;
\gamma_{ij},\;  (i,j=1,2)$, do not depend on time variable.

Variation of the action with respect to the fields leads to the equations
of motion
\begin{equation}
\frac{1}{{\sqrt g }} D_\mu \left({\sqrt g } g^{\mu\nu} D_\nu \phi \right)
= -\frac{\phi}{|\phi|}\frac{dV}{d|\phi|},\label{scaEQ}
\end{equation}
\begin{equation}
\frac{\kappa}{2}\frac{\epsilon^{\mu\nu\rho}}{{\sqrt g } }F_{\nu\rho}
=e j^{\mu}, \label{CSeq}
\end{equation}
\begin{equation}
G_{\mu\nu}=8\pi GT_{\mu\nu} .\label{Eineq}
\end{equation}
In Eq.~(\ref{Eineq}) $T_{\mu\nu}$ denotes symmetric energy-momentum tensor
\begin{eqnarray} \label{enemom}
T_{\mu\nu}&=&\frac{1}{2} \Bigl(\; \overline{D_\mu \phi}
D_\nu \phi + \overline{D_\nu \phi} D_\mu \phi \; \Bigr)
-g_{\mu\nu}\left(\frac{1}{2}g^{\rho\sigma}\overline{D_\rho\phi}{D_\sigma}\phi
-V\right).
\end{eqnarray}
This theory possesses a local U(1) gauge symmetry so that it admits a conserved
U(1) charge
\begin{equation}
Q=\int d^2\! x\,\,\!\!\sqrt{\gamma} \; j^{0},
\end{equation}
where $\gamma=\det\gamma_{ij}$ and $j^{0}$ is time-component of the U(1) current
\begin{equation} \label{current}
j^\mu = -\frac{i}{2}\left(\, \bar{\phi} {D^\mu} \phi
-\overline{{D^\mu} \phi} \phi \,\right).
\end{equation}

We are interested in the static soliton-like excitations satisfying the
classical equations of motion (\ref{scaEQ})-(\ref{Eineq}), and particularly
the self-dual solitons supported by first-order Bogomolnyi equations.
Though it is in general difficult to  have
the expression for total energy of the system, we take it
as an Euler invariant~\cite{DJH,Hen}. From time-component of the Einstein
equations
(\ref{Eineq}), we have the energy expression in terms of matter fields
\begin{eqnarray} \label{energy}
E&=&\frac{1}{16\pi G}\int d^2\! x\,\,\!\!\sqrt{\gamma}\; { }^{2}\! R \nonumber\\
&=&\int d^2\! x\,\,\!\!\sqrt{\gamma}\left\{\frac{e^{2}}{2N^{2}}A_0^{2}|\phi|^{2}
+\frac{1}{2}\gamma^{ij}\overline{\tilde{D}_{i}\phi}\tilde{D}_{j}\phi
+V(|\phi|)-\frac{3}{32\pi G}N^{2} K^{2}\right\},
\end{eqnarray}
where ${ }^{2}\! R$ is two dimensional scalar curvature, $\tilde{D_i}\phi
=(\partial_i -ieA_{i}+ieK_{i}A_0)\phi$,
$\displaystyle{K=\frac{\epsilon^{ij}}{\sqrt{\gamma}}\partial_i K_{j}}$, and
$\epsilon^{ij}$ is two
dimensional Levi-Civita tensor density of $\epsilon^{12}=\epsilon_{12}=1$.
Suppose that the potential $V$ is made of both $G$-independent $U$ and 
$G$-dependent $W$ such as $\displaystyle{V=\frac{1}{2}U^{2}+W}$. We can 
rearrange the terms in the integrand of right-hand side of Eq.~(\ref{energy})
under an assumption $N(x^{i})=1$ as follows
\begin{eqnarray} \label{BOG}
E
&=&\int d^2\! x\,\,\!\!\sqrt{\gamma}\,\Biggl\{ \frac{e^{2}}{2}|\phi|^{2}\biggl(
A_0\pm\frac{U}{e|\phi|}\biggr)^{2}\nonumber\\
&&\hspace{2cm}+\frac{1}{4}\gamma^{ij}(\overline{\tilde{D}_{i}\phi
\mp i\sqrt{\gamma}\epsilon_{ik}\gamma^{kl}\tilde{D}_{l}\phi})
(\tilde{D}_{j}\phi\mp i\sqrt{\gamma}\epsilon_{jm}\gamma^{mn}\tilde{D}_{n}\phi)
\nonumber\\
&&\hspace{2cm}\mp\frac{e}{2}(|\phi|^{2}-v^{2})\biggl[\frac{\epsilon^{ij}}{\sqrt{\gamma}}
\partial_i(A_{j}-K_{j}A_0)+KA_0+\frac{e^{2}}{\kappa}A_0|\phi|^{2}\biggr]\nonumber\\
&&\hspace{2cm}\mp eA_0|\phi|\biggl[
U-\frac{e^{2}}{2\kappa}|\phi|(|\phi|^{2}-v^{2})\biggr]\label{Rearr}\\
&&\hspace{2cm}+\biggl[
W+\frac{ev^{2}}{2}K\Bigl(\pm
\frac{A_0}{v^{2}}(|\phi|^{2}-v^{2})+\frac{ev^{2}}{2\kappa}C\Bigr)
-\frac{3}{32\pi G}K^{2}\biggr]\Biggr\}\nonumber\\
&&\pm \frac{ev^{2}}{2}\Phi\nonumber\\
&&+\int d^2\! x\,\,\,\partial_i\!\biggl[\epsilon^{ij}\Bigl(\pm\frac{ev^{2}}{2}K_{j}(A_0\mp
\frac{ev^{2}}{2\kappa}C)
\pm\frac{i}{4}(\bar{\phi}\tilde{D}_{j}\phi-\phi\overline{\tilde{D}_{j}\phi})
\Bigr)\biggr],\nonumber
\end{eqnarray}
where $\Phi$ is the magnetic flux defined by
\begin{equation}\label{mf}
\Phi=-\frac{1}{2}\int d^2\! x\,\,\epsilon^{ij}F_{ij} ,
\end{equation}
and $C$ a constant introduced so as to make the surface term on the last line vanish.
Since the spinning objects specified by the nonzero angular momentum $J$ given
as
\begin{equation}\label{ang}
J=\frac{1}{8\pi G}\oint_{|\vec{x}|\rightarrow\infty}dx^{i}K_{i}
\end{equation}
are of our interest, the requirement that the surface term in the last line of
Eq.~(\ref{BOG}) has no contribution to the energy relates $C$ to the value of scalar
magnitude at spatial infinity $\displaystyle{\phi_{\infty}
=\lim_{|\vec{x}|\rightarrow\infty}\frac{|\phi|}{v}}$ such as
\begin{equation}
C=1-\phi_{\infty}^{2}.
\end{equation}
The offdiagonal metric components $K$ in Eq.~(\ref{Rearr}) are replaced
by matter fields from $0i$-components of the Einstein equations (\ref{Eineq})
\begin{equation}\label{Kij}
K=-8\pi G\kappa
\frac{(A_0^{2}-\frac{e^{2}v^{4}}{4\kappa^{2}}D)}{N^{3}},
\end{equation}
where $D$ is an integration constant.

Now let us choose a scalar potential to achieve a Bogomolnyi-type bound. 
Specifically, $U$ is chosen to make the square bracket on the fourth 
line of Eq.~(\ref{BOG}) zero, and thereby the condition, that  
the first line of Eq.~(\ref{BOG}) should vanish, forces 
the auxiliary field $A_0$ to be
\begin{equation}
A_0=\mp\frac{U}{e|\phi|}=\mp\frac{e}{2\kappa}(|\phi|^{2}-v^{2}).
\end{equation}
The third line of Eq.~(\ref{BOG}) does not contribute to the energy because of 
the Gauss' law given by time-component of the Chern-Simons equations 
(\ref{CSeq}).
We read $G$-dependent terms of the scalar potential $W$ as an eighth-order
term from the fifth line of Eq.~(\ref{BOG}).
Hence, for the configurations to satisfy the first-order equation
\begin{equation}\label{BogAi}
\tilde{D}_{i}\phi\mp i\sqrt{\gamma}\epsilon^{ij}\gamma^{jk}\tilde{D}_{k}\phi=0,
\end{equation}
the energy is proportional to the magnetic flux and the
Bogomolnyi-type bound is saturated if such solitons carry magnetic flux.
The Gauss' law says 
that these flux-carrying solitons are charged objects and nontopological 
solitons can also be produced when the meson mass in the symmetric phase of the 
theory is not larger than $e^{2}v^{2}/2\kappa$. The remaining equations 
are $ij$-components of the Einstein equations (\ref{Eineq}) summarized
as follows
\begin{eqnarray}
\lefteqn{\frac{1}{8\pi GN}(\gamma^{ij}\nabla^{2}-\nabla^{i}\nabla^{j})N}
\nonumber\\
&=&\frac{e^{2}}{2}\gamma^{ij}|\phi|^{2}
\bigg(\frac{A_{0}}{N}-\frac{U}{e|\phi|}\bigg)
\bigg(\frac{A_{0}}{N}+\frac{U}{e|\phi|}\bigg)
-\gamma^{ij}\bigg(W+\frac{1}{32\pi G}N^{2}K\bigg) \nonumber\\
&&+\frac{1}{8}\Bigg\{\bigg[(\overline{\tilde{D}^{i}\phi\mp
i\frac{\epsilon^{ik}}{\sqrt{\gamma}}\gamma_{kl}\tilde{D}^{l}\phi})
(\tilde{D}^{j}\phi\pm i\frac{\epsilon^{jm}}{\sqrt{\gamma}}\gamma_{mn}
\tilde{D}^{n}\phi)\nonumber\\
&&\hspace{11mm}+(\overline{\tilde{D}^{j}\phi\pm
i\frac{\epsilon^{jk}}{\sqrt{\gamma}}\gamma_{kl}\tilde{D}^{l}\phi})
(\tilde{D}^{i}\phi\mp i\frac{\epsilon^{im}}{\sqrt{\gamma}}\gamma_{mn}
\tilde{D}^{n}\phi)\bigg]\nonumber\\
&&\hspace{7mm}+\bigg[(\overline{\tilde{D}^{i}\phi\pm
i\frac{\epsilon^{ik}}{\sqrt{\gamma}}\gamma_{kl}\tilde{D}^{l}\phi})
(\tilde{D}^{j}\phi\mp i\frac{\epsilon^{jm}}{\sqrt{\gamma}}\gamma_{mn}
\tilde{D}^{n}\phi)\nonumber\\
&&\hspace{11mm}+(\overline{\tilde{D}^{j}\phi\mp
i\frac{\epsilon^{jk}}{\sqrt{\gamma}}\gamma_{kl}\tilde{D}^{l}\phi})
(\tilde{D}^{i}\phi\pm i\frac{\epsilon^{im}}{\sqrt{\gamma}}\gamma_{mn}
\tilde{D}^{n}\phi)\bigg]\Bigg\}.
\end{eqnarray}
Note that any self-dual soliton solution of the Bogomolnyi equations with $N=1$
automatically satisfies the above Einstein equations.

{}From now on we fix the conformal gauge for $\gamma_{ij}$
\begin{equation}\label{confo}
\gamma_{ij}=\delta_{ij}b(x^{i}),
\end{equation}
and the Coulomb gauge for $K^{i}$, $\nabla_{i}K^{i}=0$, which makes
$K^{i}$ be expressed as follows
\begin{equation}\label{offdi}
K^{i}=-\frac{\kappa}{e^{2}v^{2}}\frac{\epsilon^{ij}}{\sqrt{\gamma}}\partial_j
\ln\psi .
\end{equation}
Substituting Eqs.~(\ref{confo})-(\ref{offdi}) into 00-component of the Einstein
equations (\ref{Eineq})
and replacing the gauge field to the scalar field by use of the
Bogomolnyi equation (\ref{BogAi})
\begin{equation}
A_{i}-K_{i}A_0=\frac{1}{e}(\partial_i\Omega\mp\epsilon^{ij}\partial_j\ln|\phi|),
\end{equation}
we obtain an expression of the metric function $b$ as
\begin{equation} \label{bbb}
b=e^{h(\tilde{z})+\bar{h}(\bar{\tilde{z}})}\left(\frac{f^{2}e^{-(f^{2}-1)}
\psi^{1-\phi_{\infty}^{2}}}{\prod_{p=1}^{n}
|\tilde{z}-\tilde{z}_{p}|^{2}}\right)^{\tilde{G}}.
\end{equation}
In the above $h(\tilde{z})$ $(\bar{h}(\bar{\tilde{z}}))$ is a holomorphic
(an anti-holomorphic) function
and the variables with tilde are dimensionless quantities
\begin{equation}
\tilde{z}=\tilde{x}^{1}+i\tilde{x}^{2}=\frac{e^{2}v^{2}}{|\kappa|}(x^{1}+ix^{2}),\;\;
f=\frac{|\phi|}{v},\;\;\tilde{G}=4\pi Gv^{2}.
\end{equation}

If we eliminate the gauge field and the metric $b$ by use of
Eqs.~(\ref{BogAi})-(\ref{bbb}), we have a Bogomolnyi equation from the
Gauss' law which is the time-component of the Chern-Simons equations
(\ref{CSeq}) for the gauge field
\begin{eqnarray} \label{Bogm}
\tilde{\partial}^{2}\ln f^{2}&=&
e^{h(\tilde{z})+\bar{h}(\bar{\tilde{z}})}\left(\frac{f^{2}e^{-(f^{2}-1)}
\psi^{1-\phi_{\infty}^{2}}}{\prod_{p=1}^{n}
|\tilde{z}-\tilde{z}_{p}|^{2}}\right)^{\tilde{G}}
(f^{2}-1)\nonumber\\
&&\times\left[f^{2}-\frac{\tilde{G}}{2}\left(f^{4}-2f^{2}+\phi^{2}_{\infty}\right)
\right]\mp 2\epsilon^{ij}\tilde{\partial_i}\tilde{\partial_j}\Omega,
\end{eqnarray}
and the equation for $\psi$ from Eq.~(\ref{Kij})
\begin{equation}\label{eqK}
\tilde{\partial}^{2}\ln\psi=-\frac{\tilde{G}}{2}
e^{h(\tilde{z})+\bar{h}(\bar{\tilde{z}})}\left(\frac{f^{2}e^{-(f^{2}-1)}
\psi^{1-\phi_{\infty}^{2}}}{\prod_{p=1}^{n}
|\tilde{z}-\tilde{z}_{p}|^{2}}\right)^{\tilde{G}}
\left(f^{4}-2f^{2}+\phi^{2}_{\infty}\right),
\end{equation}
where $\tilde{\partial}^{2}$ is flat-space Laplacian, i.e., we will use
$\tilde{\partial}^{2}\equiv\tilde{\partial}_{i}\tilde{\partial}_{i}
=\Delta$.

\begin{figure}

\setlength{\unitlength}{0.1bp}
\begin{picture}(3600,2807)(0,0)
\put(1390,2000){\makebox(0,0)[r]{$\phi_{\infty}=\sqrt{1+1/2\pi Gv^{2}}$}}
\put(1106,2200){\makebox(0,0)[r]{$\phi_{\infty}=0$}}
\put(1106,2400){\makebox(0,0)[r]{$\phi_{\infty}=1$}}
\put(20,1000){\makebox(0,0)[r]{{\Large 0}}}
\put(1980,-100){\makebox(0,0)[r]{{\Large $|\phi|$}}}
\put(2020,1100){\makebox(0,0)[r]{{\Large $v$}}}
\put(0,1400){\makebox(0,0)[r]{{\Large $V(|\phi|)$}}}
\end{picture}

\vspace{3mm}
\caption{Schematic shapes of the scalar potential $V(|\phi|)$ for
$e^{4}/8\kappa^{2}=1$ and $\pi Gv^{2}=1$. The solid, dashed, and dotted lines
correspond to $\phi_{\infty}=1$, 0, and
$\sqrt{1+1/2\pi Gv^{2}}$.}
\label{fig1}
\end{figure}
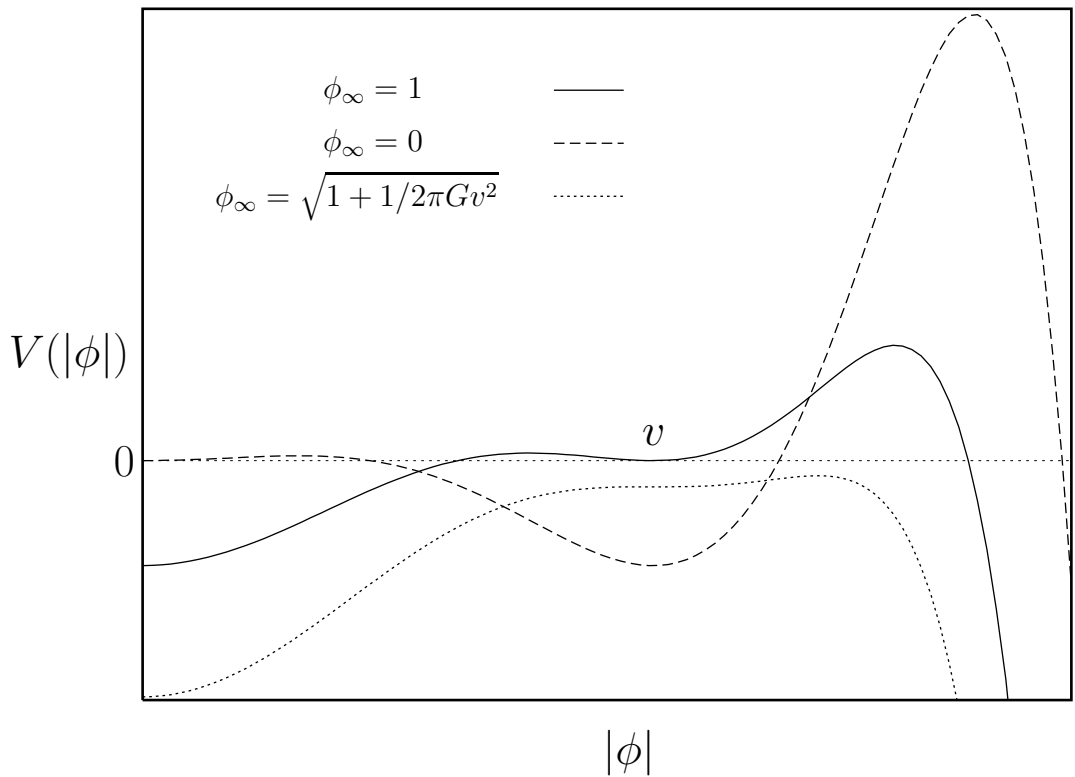

The undetermined constant $D$ in Eq.~(\ref{Kij}) is fixed by the requirement
that the above equations (\ref{Bogm})-(\ref{eqK}) reproduce the scalar equation
(\ref{scaEQ}), $D=1-\phi_{\infty}^{2}$. In synthesis, the scalar potential
becomes
\begin{equation}\label{pot}
V(|\phi|)=\frac{e^{4}}{8\kappa^{2}}\Bigl[|\phi|^{2}(|\phi|^{2}-v^{2})^{2}-\pi G
(|\phi|^{4}-2v^{2}|\phi|^{2}+v^{4}\phi_{\infty}^{2})^{2}\Bigr].
\end{equation}
Though the Bogomolnyi limit of a theory selects usually a unique scalar
potential in curved spacetime~\cite{CG},
shape of the scalar potential (\ref{pot}) depends on the boundary value
of the scalar field $\phi_{\infty}$ (see also Fig.~\ref{fig1}).
This potential includes $G$-dependent eighth-order potential,
and it is not bounded below since the coefficient of the eighth-order term is
negative for positive Newton's constant $G$ as shown in Fig.~\ref{fig1}.
However the energy in order to support flux-carrying solitons of the Bogomolnyi
equations is positive definite. Inserting the metric (\ref{bbb}) into the
Euler characteristic (\ref{energy}), we have
\begin{eqnarray}\label{Eul}
\lefteqn{\frac{1}{4\pi Gv^{2}}\int d^2\! x\,\,\!\!\sqrt{\gamma}\;{}^{2}\! R} \nonumber\\
&=&\frac{1}{4\pi Gv^{2}}\left\{\int d^2\! x\,\, \partial^2 \ln\prod_{p=1}^{n}|z-z_{p}|^{2}
-\int d^2\! x\,\, \partial^2 \ln |\phi|^2 + \int d^2\! x\,\,
\partial^2\frac{|\phi|^2}{v^{2}} \right. \nonumber\\
&& \hspace{18mm}-(1-\phi^{2}_{\infty})\int d^2\! x\,\,\partial^{2}\ln\psi
\Bigg\} .
\end{eqnarray}
{}From the above expression there can exist
contributions from the second and fourth terms besides the first term
for $n\ne 0$ if the Bogomolnyi
equations (\ref{Bogm})-(\ref{eqK}) contain the finite energy solution which behaves as
$|\phi| \sim |x^{i}|^{-\varepsilon} \; (\varepsilon > 0) $
for large $|\vec{x}|$, and $\psi\sim |x^{i}|^{-2\alpha}\;(\alpha$ is an
arbitrary number) when $\phi_{\infty}\neq1$. We shall present
the detailed analysis for the existence of such solutions in section 4.

Looking at shapes of the scalar potential in Fig.~\ref{fig1}, one
may raise an intriguing question in connection with local
supersymmetry in 2+1 dimensions. According to Ref.~\cite{BBS}, a
well-known argument in 2+1 dimensions is that local supersymmetry
can ensure the vanishing of the cosmological constant without
requiring the equality of boson and fermion masses. Suppose that
the above theory is bosonic part of a presumed supergravity theory
as has been done for $N=2$ supersymmetric theory of self-dual
Chern-Simons Higgs model in flat spacetime~\cite{LLW}. It is
questionable that whether or not the supergravity version of our
model satisfies the above criterion because of the following
reasons. Although we let the cosmological constant vanish from
the beginning, the scalar potential (\ref{pot}) is unbounded
below and has negative local minima. Specifically, degeneracy
between the symmetric vacuum, $|\phi|=0$, and the broken vacua,
$|\phi|=v$, disappears because of $G$-dependent terms, and this
phenomenon is guaranteed by global supersymmetry~\cite{LLW} (see
Fig.~\ref{fig1}). In more detail, the value of the scalar
potential for the boundary values of soliton configurations are
calculated as follows;
\begin{equation}\label{bdpot}
V(|\phi|=v\phi_{\infty})=\frac{\pi}{8}\frac{Ge^{4}v^{8}}{\kappa^{2}}
\phi_{\infty}^{2}(\phi^{2}_{\infty}-1)^{2}\left(\phi^{2}_{\infty}
-\frac{1}{\pi Gv^{2}}\right).
\end{equation}
If the boundary value of the scalar field is different from
$0, 1, 1/\sqrt{\pi Gv^{2}}$, then $V(\phi_{\infty})\ne 0$ and
we may suspect that our model is not embedded in the
2+1-dimensional supergravity models claimed in Ref.~\cite{BBS}.
In section 4, we will show that all possible regular self-dual
solitons with rotational symmetry have boundary conditions $\phi_{\infty}=0$ or
1 so that the model of our consideration shares properties of the vanishing
cosmological constant, i.e.,
$V(|\phi|=0;\phi_{\infty}=0)=V(|\phi|=v;\phi_{\infty}=1)=0$. One more comment
should be placed: For both cases to support regular self-dual solitons
(nonperturbative spectra), the potential change due to $G$-dependent terms
does not affect the mass spectra of perturbative gauge and scalar bosons.
Therefore, equality of the gauge boson mass and the Higgs mass in broken phase
$(\phi_{\infty}=1)$, still holds in curved spacetime, i.e.,
$m_{\rm gauge}=m_{\rm Higgs}=e^{2}v^{2}/|\kappa|$.
The above properties imply that one of our present systems classified by
an undetermined parameter $\phi_{\infty}$ may be obtained as the bosonic
sector of some supergravity models.

\setcounter{equation}{0}
\section{Bogomolnyi Bound in Dual Formulation}
We will construct duality transformed theory of our interest
and see how the solitons arise in this framework. Let us work
in the path integral formalism
\begin{equation} \label{Dual}
Z=\int [dg_{\mu\nu} ][dA_{\mu}][|\phi| d|\phi| ][d\Omega ]e^{iS},
\end{equation}
where the action $S$ was defined in Eq.~(\ref{action}). The interaction between
the scalar and gauge fields is linearized by introducing an auxiliary field
$C_{\mu}$ as follows
\begin{eqnarray} \label{Lin}
\lefteqn{\exp \left\{ i\int d^3\! x\,\,\!{\sqrt g }\,
\frac{1}{2}g^{\mu\nu}|\phi|^{2}(\partial_\mu\Omega-eA_{\mu})
   ({\partial_\nu}\Omega-eA_{\nu})\right\} }\nonumber\\
&=&\int [dC_{\mu}]\prod_{x}\frac{g^{\frac{1}{4}}}{|\phi|^{3}}\exp i\int d^3\!
x\,\,
\!{\sqrt g }\left\{ -\frac{g^{\mu\nu}}{2|\phi|^{2}}
C_{\mu}C_{\nu}+g^{\mu\nu} C_{\mu}({\partial_\nu}\Omega -eA_{\nu} )\right\}.
\end{eqnarray}
An auxiliary vector field $C_{\mu}$ is classically identified with the U(1)
current by constraint equation.
Contribution of the scalar phase $\Omega(x)$ is divided into two: one from
topological excitations described by a multi-valued function $\Theta(x)$
and the other from fluctuations around superselected topological sector
expressed by a single-valued function $\eta(x)$. Then the path integral measure
for the scalar phase becomes
\begin{equation}
[d\Omega]=[d\Theta][d\eta],
\end{equation}
and $\eta$-integration explains conservation of the U(1) current in the path
integral formalism
\begin{equation} \label{CuCon}
\int [d\eta] \exp \left\{ i\int d^3\! x\,\,\!{\sqrt g }\,g^{\mu\nu}
C_{\mu}{\partial_\nu}\eta
\right\}\approx \frac{1}{{\sqrt g }}\delta(\nabla_{\mu} C^{\mu}).
\end{equation}
Dual vector field $H_{\mu}$ is introduced such as
\begin{equation} \label{Meas}
\int [dC_{\mu}]\frac{1}{{\sqrt g }}\delta (\nabla_{\mu}C^{\mu})\cdots
=\int [dH_{\mu}][dC_{\mu}]\delta({\sqrt g }\, C^{\mu}-\frac{\kappa}{e}
\epsilon^{\mu\nu\rho\sigma}{\partial_\nu} H_{\rho})\cdots\, ,
\end{equation}
where the scale $\kappa/e$ is introduced for later convenience.
The action at this stage is quadratic in both $C_{\mu}$ and $A_\mu$ so that
the path integrals are Gaussian type. Integrating both $A_\mu$ and $C_{\mu}$
in a closed form, we obtain duality-transformed theory;
\begin{eqnarray}
Z&=&\int [g^{\frac{3}{4}}dg_{\mu\nu}][d H_{\mu}][|\phi|^{-2}d|\phi|][d\Theta ]
\nonumber\\
&&\exp i\int d^3\! x\,\,\!\!{\sqrt g } \left\{-\frac{1}{16\pi G}R-\frac{\kappa^{2}
}{4e^{2}|\phi|^{2}}g^{\mu\rho}g^{\nu\sigma} H_{\mu\nu}H_{\rho\sigma}-\frac{\kappa}{2}
\frac{\epsilon^{\mu\nu\rho}}{\sqrt{g}}H_{\mu}{\partial_\nu} H_{\rho}\right. \nonumber\\
&&\hspace{2.8cm}\left.
+\frac{\kappa}{2e}\frac{\epsilon^{\mu\nu\rho}}{{\sqrt g
}}H_{\mu\nu}\partial_\rho\Theta +\frac{1}{2}
g^{\mu\nu}\partial_\mu|\phi|{\partial_\nu}|\phi| -V(|\phi|) \right\},
\label{dualz}
\end{eqnarray}
where $H_{\mu\nu}=\partial_\mu H_{\nu}-{\partial_\nu} H_{\mu}$.

Equations of motion read
\begin{equation}\label{dsc}
\frac{1}{\sqrt{g}}\partial_\mu(\sqrt{g}g^{\mu\nu}{\partial_\nu}|\phi|)=-\frac{dV}{d|\phi|}+\frac{\kappa^{2}}{2e^{2}|\phi|^{3}}
g^{\mu\rho}g^{\nu\sigma} H_{\mu\nu}H_{\rho\sigma},
\end{equation}
\begin{equation}\label{dcseq}
\frac{1}{\sqrt{g}}{\partial_\nu}\left(\sqrt{g}\,\frac{H^{\mu\nu}}{|\phi|^{2}}\right)+\frac{e^{2}}{2\kappa}
\frac{\epsilon^{\mu\nu\rho\sigma}}{{\sqrt g
}}H_{\nu\rho}=\frac{e}{\kappa}\frac{\epsilon^{\mu\nu\rho}}{{\sqrt g
}}{\partial_\nu}\partial_\rho
\Theta,
\end{equation}
\begin{equation}\label{dei}
G_{\mu\nu}=8\pi GT_{\mu\nu}^{D},
\end{equation}
where the energy-momentum tensor is
\begin{eqnarray}
T^{D}_{\mu\nu}&=&\partial_\mu|\phi|{\partial_\nu}|\phi|-\frac{\kappa^{2}}{e^{2}|\phi|^{2}}
g^{\rho\sigma} H_{\mu\rho}H_{\nu\sigma}\\
&&-g_{\mu\nu}\left\{\frac{1}{2}g^{\alpha\beta}\partial_\alpha|\phi|
\partial_\beta|\phi|-\frac{\kappa^{2}}{4e^{2}|\phi|^{2}}g^{\alpha\beta}g^{\gamma\delta} H_{\alpha\gamma}
H_{\beta\delta}-V\right\}\nonumber.
\end{eqnarray}
In the Higgs phase with $|\phi|=v$, the interaction term between the Higgs
field and the dual gauge field turns out to be Maxwell term so that the
Chern-Simons Higgs theory is transformed to topologically massive gauge theory
and thereby both
theories support an odd-parity helicity-one photon of mass
$e^{2}v^{2}/|\kappa|$~\cite{DJ}. Even though the action
in the duality-transformed theory (\ref{dualz}) looks inappropriate to
describe the symmetric phase
of Chern-Simons scalar electrodynamics because of the singular factor
$1/|\phi|^{2}$ in both the action and the path integral measure,
this term plays an important role to produce nonperturbative spectra, e.g.,
nontopological solitons. For this, let us look at the Gauss' law in flat
spacetime
\begin{equation}\label{dga}
\partial_i\left(\frac{E^{i}}{|\phi|^{2}}\right)+\frac{e^{2}}{\kappa}B=\pm\frac{2\pi e}{\kappa}
\sum_{p=1}^{n}\delta^{(2)}(x^{i}-x_{p}^{i}).
\end{equation}
Since the electric field term in the left-hand side of Eq.~(\ref{dga}) has no
long-range contribution in the Higgs phase, the magnetic flux must be quantized.
If there exists a soliton of which the scalar field behaves like
$|\phi|\sim|x^{i}|^{-\varepsilon}$ $ (\varepsilon >0)$ for large $|x^{i}|$
in addition to the discrete magnetic flux,
the magnetic flux has additional contribution to compete with that
of the electric field term (\ref{dga}), which is not necessarily
discrete. It is indeed the case of the Chern-Simons Higgs theory~\cite{JLW}.
Since the gauge coupling $e$ is inversely coupled to the Maxwell-like term,
which describes interaction between the scalar and
gauge fields, the strong coupling expansion is achievable when the
nonpolynomial Higgs interaction is neglected~\cite{ES}. Note that, though the
classical gravity is not affected by the duality transformation, path integral
measure for the gravitational field contains a nontrivial Jacobian factor
which comes from gauge dynamics~\cite{KK}.

{}From now on we explore the static solitons given by solutions of the classical
equations of motion. The duality transformation makes the system
complicate by changing the first-order Chern-Simons equation (\ref{CSeq}) to
the second-order equation (\ref{dcseq}), however Eq.~(\ref{dcseq}) is reduced to
the first-order Chern-Simons equation if we replace the gauge
field $A_\mu$ by the dual gauge field $H_{\mu}$. It is due to the fact that
a relation between the auxiliary field $C_{\mu}$ and the dual gauge field
$H_{\mu}$ in the right-hand side of Eq.~(\ref{Meas}) is the form of
classical Chern-Simons equation. Instead of solving the Euler-Lagrange
equations (\ref{dsc})-(\ref{dei}), we derive a Bogomolnyi-type bound of
the duality-transformed Einstein Chern-Simons Higgs
theory again. Similar to the previous section, we define the energy
as Euler invariant and assume the 00-component of stationary metric $N^{2}$
to be 1. Reshuffle of the terms of the energy gives us an expression of the
energy such as
\begin{eqnarray}\label{dbog}
E_{D}&=&\int d^2\! x\,\,\!\!\sqrt{\gamma}\,\Biggl\{\frac{\kappa^{2}}{2e^{2}|\phi|^{2}}\gamma^{ij}
\partial_i\Bigl(H_{0}\pm\frac{e}{2\kappa}(|\phi|^{2}-v^{2})\Bigr)\partial_j
\Bigl(H_{0}\pm\frac{e}{2\kappa}(|\phi|^{2}-v^{2})\Bigr)
\nonumber\\
&&\hspace{20mm}+\frac{1}{2}\Bigl(\frac{\kappa}{e|\phi|}\tilde{H}\mp
U\Bigr)^{2}\pm\frac{\kappa}{2e}|\phi|^{2}\biggl[\Bigl[\frac{1}{\sqrt{\gamma}}\partial_i\Bigl(\sqrt{\gamma}
\gamma^{ij}\frac{1}{|\phi|^{2}}(\partial_j H_{0}+\gamma^{kl}K_{l}\tilde{H}_{jk})\Bigr)
\nonumber\\
&&\hspace{20mm}+\frac{e^{2}}{2\kappa}\frac{\epsilon^{ij}}{\sqrt{\gamma}}H_{ij}-\frac{e}{\kappa}
\frac{\epsilon^{ij}}{\sqrt{\gamma}}\partial_i\partial_j\Theta\Bigr]
-\Bigl[\frac{\epsilon^{ij}}{\sqrt{\gamma}}\partial_i\Bigl(K_{j}(\frac{\kappa}{e}H_{0}+\frac{\tilde{H}}{|\phi|^{2}})\Bigr)\Bigr]\biggr]\nonumber\\
&&\hspace{20mm}
\pm\frac{\kappa}{e|\phi|}\tilde{H}\Bigl( U-\frac{e^{2}}{2\kappa}
|\phi|(|\phi|^{2}-v^{2})\Bigr)\\
&&\hspace{20mm}+\biggl(W+\frac{e}{2}K\Bigl(\pm(|\phi|^{2}-v^{2})H_{0}+
\frac{ev^{4}}{2\kappa}
C_{D}\Bigr)-\frac{3}{32\pi G}K^{2}\biggr)\Biggr\}\nonumber\\
&&\mp\frac{ev^{2}}{4}\int d^2\! x\,\,\epsilon^{ij} H_{ij}\nonumber\\
&&+\int d^2\! x\,\,\partial_i\biggl[\,\pm\frac{ev^{2}}{2}\epsilon^{ij} K_{j}\Bigl(H_{0}\mp\frac{ev^{2}}{2\kappa}C_{D}
\Bigr)\partial_i\mp\frac{\kappa}{2e}\sqrt{\gamma}\gamma^{ij}\partial_j H_{0}
\biggr]\nonumber,
\end{eqnarray}
where $\displaystyle{\tilde{H}=\frac{\epsilon^{ij}}{2\sqrt{\gamma}}\tilde{H}_{ij}=
\frac{\epsilon^{ij}}{2\sqrt{\gamma}}(H_{ij}+K_{i}\partial_j H_{0}-K_{j}\partial_i H_{0})}$, and $C_{D}$
is an integration constant.

Similar to the procedure done in the section 2, we obtain the Bogomolnyi-type
bound for the duality-transformed Einstein Chern-Simons Higgs theory.
Square brackets in the second and third
lines in Eq.~(\ref{dbog}) vanish by use of the equations for the gauge fields
(\ref{dcseq}). Let us assume that configurations satisfy
\begin{equation}
H_{0}=\mp\frac{e}{2\kappa}(|\phi|^{2}-v^{2}).
\end{equation}
Then we can fix $C_{D}$ as $1-\phi^{2}_{\infty}$ from the condition that the
objects of our interest have spin $J$ and the boundary terms in the last line
of Eq.~(\ref{dbog}) vanish. With the aid of the solutions of $0i$-components
of the Einstein equations, we choose a specific form of the scalar potential
to let the terms in the fourth and fifth lines
be zero. Hence the Bogomolnyi-type bound is attained for the solutions of the
following equation
\begin{equation}\label{db}
\tilde{H}=\pm\frac{e|\phi|}{\kappa}U=\pm\frac{e^{3}}{2\kappa}|\phi|^{2}(|\phi|^{2}-v^{2}).
\end{equation}
Replacing the gauge and gravitational fields in Eq.~(\ref{db}) by use
of the relations obtained above, we have the same Bogomolnyi equations in
Eq.~(\ref{Bogm}) and Eq.~(\ref{eqK}). Among the remaining equations,
solutions of the Bogomolnyi equation
satisfy $ij$-components of the Einstein equations (\ref{dei}) and
reproducibility of the scalar equation (\ref{dsc}) determines the integration
constant introduced by $0i$-component of the Einstein equations.

\setcounter{equation}{0}
\section{Soliton Solutions}
In this section we explore soliton solutions by examining the
equations of motion, i.e., one is Bogomolnyi equation
(\ref{Bogm}) and the other is the remaining Einstein equation
(\ref{eqK}). We restrict our interest to  rotationally symmetric
configurations  satisfying the following three conditions; \par
\noindent 1. nonsingular solutions of equations of motion,\par
\noindent 2. solutions of finite energy which is proportional to
the magnetic flux, and \par \noindent 3. the underlying manifold
defined by the solution does not include curvature singularity.

The stationary metric ~(\ref{metR}) compatible with rotationally
symmetric solutions becomes
\begin{equation} \label{rotmet}
ds^{2}=\left(dt+\frac{\kappa}{e^{2}v^{2}}\frac{d\ln\psi(r)}{dr}rd\theta\right)^{2}
-\left(\frac{\kappa}{e^{2}v^{2}}\right)^{2}b(r)(dr^{2}+r^{2}d\theta^{2}),
\end{equation}
where $r=\sqrt{\tilde{x}^{i}\tilde{x}^{i}}$. With the help of a
gauge transformation, ansatz for the scalar field is
\begin{equation}\label{ini}
\phi=vf(r)e^{-in\theta},
\end{equation}
and thereby a component of the  metric $b$ ~(\ref{bbb}) becomes
\begin{equation}\label{bbr}
b(r)=F\left[\frac{f^{2}e^{-(f^{2}-1)}\psi^{1-\phi^{2}_{\infty}}}{r^{2n}}
\right]^{\tilde{G}},
\end{equation}
where $F$ is an undetermined constant due to the harmonic function
$h(\tilde{z})$. Under this $(r,\theta)$-coordinate, the equations
of motion (\ref{Bogm}) - (\ref{eqK}) become
\begin{eqnarray}
\frac{1}{r}\frac{d}{dr}r\frac{d}{dr}\ln\frac{f^{2}}{r^{2n}}&=&
-\frac{F\tilde{G}}{2}\left[\frac{f^{2}e^{-(f^{2}-1)}
\psi^{1-\phi^{2}_{\infty}}}{r^{2n}}\right]^{\tilde{G}}
(f^{2}-1)\Bigl[f^{4}-2(1+\frac{1}{\tilde{G}})f^{2}+\phi_{\infty}^{2}
\Bigr],\label{fff}\\
\frac{1}{r}\frac{d}{dr}r\frac{d}{dr}\ln\psi&=&-\frac{F\tilde{G}}{2}
\left[\frac{f^{2}e^{-(f^{2}-1)}\psi^{1-\phi^{2}_{\infty}}}{r^{2n}}\right]^{\tilde{G}}
(f^{4}-2f^{2}+\phi_{\infty}^{2}).\label{ppp}
\end{eqnarray}
 To obtain nonsingular solution at the origin, Eq.~(\ref{ini})
fixes the boundary condition of the scalar field at $r=0$
\begin{equation}
nf(r=0)=0
\end{equation}
which implies the leading behavior of scalar field for small $r$
\begin{equation}\label{f0}
f\sim f_{0}r^{n},
\end{equation}
and Eq.~(\ref{ppp}) gives the boundary condition of $\psi$ at $r=0$
\begin{equation}
\psi(r=0)=\psi_{0},
\end{equation}
where $\psi_{0}$ is an arbitrary positive constant.

Rewriting the magnetic flux ~(\ref{mf}) in terms of $f$ and $\psi$
\begin{equation}\label{mfr}
\Phi=\pm\frac{2\pi}{e}\left.\biggl[n-r\frac{d\ln f}{dr}
-\frac{1}{2}(1-f^{2})r\frac{d\ln\psi}{dr}\biggr]\right|_{r\rightarrow\infty}
\end{equation}
and the angular momentum ~(\ref{ang}) as
\begin{eqnarray}
J&=&-\frac{\pi\kappa}{2e^{2}}\int^{\infty}_{0}\! dr\,r\, F
\left(\frac{f^{2}e^{-(f^{2}-1)}\psi^{1-\phi^{2}_{\infty}}}{r^{2n}}\right)^{\tilde{G}}
(f^{4}-2f^{2}+\phi^{2}_{\infty}) \label{an1}\\
&=&\left.-\frac{\pi\kappa}{e^{2}\tilde{G}}r\frac{d}{dr}\ln\psi
\right|_{r=\infty},
\label{an2}
\end{eqnarray}
we notice that, from Eq.~(\ref{an2}), spinning objects have long
range term of $\psi$ for large $r$, specifically $\psi\sim
r^{-2\alpha}$, where $\alpha$ is an arbitrary number and it also
contributes to the magnetic flux ~(\ref{mfr}) if
$\phi_{\infty}\neq1$. The energy ~(\ref{energy}) can also be
expressed in terms of $f$ and $\psi$ such as
\begin{eqnarray}\label{eee}
E&=&\pi v^{2}\int_{0}^{\infty}\! dr\,
r\,\Biggl\{\biggl(\frac{df}{dr}\biggr)^{2} +\frac{F}{4}
\left(\frac{f^{2}e^{-(f^{2}-1)}\psi^{\phi^{2}_{\infty}-1}}{r^{2n}}\right)^{\tilde{G}}
\Bigl[
f^{2}(f^{2}-1)^{2}-\frac{\tilde{G}}{2}(f^{4}-2f^{2}+\phi^{2}_{\infty})^{2}\Bigr]
\Biggr\}\nonumber\\
&=&\frac{v^{2}}{2}|e\Phi|.
\end{eqnarray}
Eqs.~(\ref{eee}) -~(\ref{an1}) give  possible boundary conditions
of the  scalar amplitude  at $r=\infty$ for a given $\alpha$
\begin{equation}
\frac{|\phi|}{v}=\left\{ \begin{array}{ll}
\mbox{0 or 1}, & \mbox{when $\tilde{G}[n+\alpha(1-\phi^{2}_{\infty})]\leq 1$}\\
\mbox{arbitrary}, & \mbox{when
$\tilde{G}[n+\alpha(1-\phi^{2}_{\infty})]> 1$.}
\end{array} \right.
\end{equation}

About the geometry of two dimensional space $\Sigma$ described by
the $\gamma_{ij}$ part of Eq.~(\ref{metR}) (or by the $b(r)$ of
Eq.~(\ref{rotmet})),  various manifolds are characterized by  the
area of space defined by $\displaystyle{{\cal A}=
2\pi\int_{0}^{\infty} \! dr\, r\, b(r)}$. Suppose that the
asymptotic behavior of the scalar amplitude as $|\phi|/v\sim
r^{-\varepsilon}$ where $\varepsilon$ is related to $\alpha$, the
above categories are distinguished  by the value of
 $\tilde{G}[n+\varepsilon+(1-\phi^{2}_{\infty})\alpha]$, i.e., whether it is less
than or equal  or greater than 1. Additional quantities
characterizing the property of $\Sigma$ are the radial distance
from the origin
$\rho(r)=(e^{2}v^{2}/|\kappa|)\int_{0}^{r}dr^{'}\sqrt{b(r^{'})}$
and the circumference $l(r)=(2\pi
e^{2}v^{2}/|\kappa|)r\sqrt{b(r)}$.
When $\varepsilon=0$, the above expression of the scalar amplitude
$| \phi|/v \sim r^{-\varepsilon}$ corresponds to a nonzero finite
$\phi_{\infty}$. Rapid decay of the scalar amplitude is depicted
as $\varepsilon \to \infty$.  So this assumption may not lose any
generality.  In this section we will obtain the rotationally
symmetric solutions  carrying magnetic flux (or charge) and spin.
Underlying spatial manifold $\Sigma$ will comprise three types;
cone, cylinder, and two sphere. We address this question
separately for these three cases.

{}From now on, let us examine precisely all possible rotationally
symmetric solutions of the Bogomolnyi equations
(\ref{fff})-(\ref{ppp}). When $\phi_{\infty}=1$, Eq.~(\ref{fff})
for $f$ is decoupled from the equation for $\psi$ ~(\ref{ppp}).
Thus it may be convenient to investigate the soliton solutions for
the cases separately by whether  $\phi_{\infty}$ is equal to one
or not.

\vspace{5mm}

\subsection*{\normalsize\bf (a) \underline{$\phi_{\infty}=1$}}
\indent\indent We consider the case $\phi_{\infty}=1$ by dividing
it into two categories: $\tilde{G}n\neq 1$ and $\tilde{G}n=1$.
When $\tilde{G}n$ is smaller than one, we introduce variables $u$
and $R$ such that
\begin{equation}\label{ufu}
u=\ln f^{2}
\end{equation}
and
\begin{equation}\label{RRR}
R=\frac{r^{1-\tilde{G}n}}{1-\tilde{G}n}.
\end{equation}
Then the Bogomolnyi equation (\ref{fff}) is rewritten as
\begin{equation} \label{New1}
\frac{d^{2}u}{dR^{2}}=-\frac{dU_{\rm
eff}}{du}-\frac{1}{R}\frac{du}{dR},
\end{equation}
where $U_{\rm eff}$ is given by
\begin{equation} \label{veff}
U_{\rm
eff}(u;\phi_{\infty}=1)=\displaystyle{-\frac{F}{\tilde{G}}\exp\Bigl[-
\tilde{G}(e^{u}-u-1)\Bigr]}(e^{u}-1)^{2}.
\end{equation}
As shown in Fig.~\ref{fig2}-(a), $U_{\rm eff}(u;\phi_{\infty}=1)$
has two local minima at
$u=\ln[(\tilde{G}+1\pm\sqrt{2\tilde{G}+1})/\tilde{G}]$ and three
maxima at $u=0,\;\pm\infty$. If we interpret $u$ as the position
of a hypothetical particle with unit mass and $R$ as time,
Eq.~(\ref{New1}) is Newton's equation for 1-dimensional motion of
the hypothetical particle moving in the potential $U_{\rm eff}$
and subject to a time-dependent friction, $-(du/dR)/R$. For
$n\neq0$ cases, the particle also receives an impact at $R=0$
from the delta function term, i.e.,
$[2n/(1-\tilde{G}n)^{1/(1-\tilde{G}n)}]
\delta(R^{1/(1-\tilde{G}n)})$ from Eq.~(\ref{Bogm}). The energy of
the hypothetical particle ${\cal E}(R)$ is defined by
\begin{equation} \label{fir}
{\cal E}(R)=\frac{1}{2}\left(\frac{du}{dR}\right)^{2}+U_{\rm
eff}(u),
\end{equation}
and it is a monotonically decreasing function of $R$ because of
the friction. When $n=0$, trivial solution $u=0$ is the unique
solution which describes two dimensional flat space. We now show
that, only when $0<\tilde{G}n<1$, there always exists a finite
energy solution whose base manifold $\Sigma$ is an asymptotic
cone.
\begin{figure}
\vspace{-8mm}

\setlength{\unitlength}{0.1bp}
\begin{picture}(3600,2807)(0,0)
\put(1850,100){\makebox(0,0){\Large $u$}}
\put(-50,1478){%
\makebox(0,0)[b]{\shortstack{\Large $U_{eff}$}}%
}
\put(1266,1200){\makebox(0,0)[r]{\large $\phi_{\infty}=1$}}
\put(1266,1000){\makebox(0,0)[r]{\large $\phi_{\infty}=0$}}
\put(1554,800){\makebox(0,0)[r]{\large $\phi_{\infty}=\sqrt{1+1/2\pi Gv^{2}}$}}
\put(2185,2293){\makebox(0,0){\Large 0}}
\put(1200,2423){\makebox(0,0){
$u=\ln[(\tilde{G}+1-\sqrt{2\tilde{G}+1})/\tilde{G}]$}}
\put(1720,2207){\makebox(0,0){$|$}}
\end{picture}

\vspace{-7mm}
\begin{center}{(a)}
\end{center}

\setlength{\unitlength}{0.1bp}
\begin{picture}(3600,2807)(0,0)
\put(2700,130){\makebox(0,0){\Large $0$}}
\put(1850,100){\makebox(0,0){\Large $\xi$}}
\put(-50,1478){%
\makebox(0,0)[b]{\shortstack{\Large $W_{eff}$}}%
}
\end{picture}

\vspace{-7mm}
\begin{center}{(b)}
\end{center}
\vspace{-4mm} \caption{(a) Typical shapes of $U_{\rm eff}$: The
solid, dashed, and dotted lines correspond to $\phi_{\infty}=1$,
0, and $\sqrt{1+1/2\pi Gv^{2}}$. (b) A typical shape of $W_{\rm
eff}$.} \label{fig2}
\end{figure}

For this we  prove that, for a suitably-chosen initial parameter
$f_{0}$, we can obtain a motion of the hypothetical particle such
that it starts at negative infinity with the initial velocity
given by Eq.~(\ref{f0}) and stops at $u=0$ at $R=\infty$.

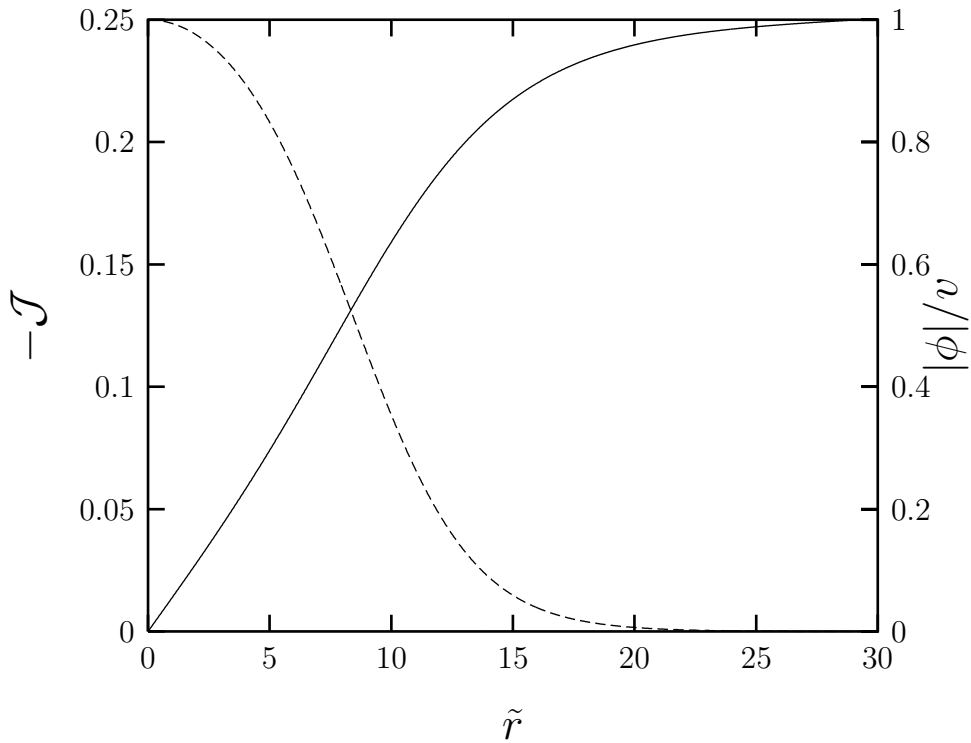
\begin{figure}

\setlength{\unitlength}{0.1bp}
\begin{picture}(3600,2807)(0,0)
\put(1875,50){\makebox(0,0){\Large $\tilde{r}$}}
\put(3550,1553){%
\makebox(0,0)[b]{\shortstack{\Large $|\phi|/v$}}%
}
\put(100,1553){%
\makebox(0,0)[b]{\shortstack{\Large ${\cal -J}$}}%
}
\put(3300,2707){\makebox(0,0)[l]{1}}
\put(3300,2246){\makebox(0,0)[l]{0.8}}
\put(3300,1784){\makebox(0,0)[l]{0.6}}
\put(3300,1323){\makebox(0,0)[l]{0.4}}
\put(3300,861){\makebox(0,0)[l]{0.2}}
\put(3300,400){\makebox(0,0)[l]{0}}
\put(3250,300){\makebox(0,0){30}}
\put(2792,300){\makebox(0,0){25}}
\put(2333,300){\makebox(0,0){20}}
\put(1875,300){\makebox(0,0){15}}
\put(1417,300){\makebox(0,0){10}}
\put(958,300){\makebox(0,0){5}}
\put(500,300){\makebox(0,0){0}}
\put(450,2707){\makebox(0,0)[r]{0.25}}
\put(450,2246){\makebox(0,0)[r]{0.2}}
\put(450,1784){\makebox(0,0)[r]{0.15}}
\put(450,1323){\makebox(0,0)[r]{0.1}}
\put(450,861){\makebox(0,0)[r]{0.05}}
\put(450,400){\makebox(0,0)[r]{0}}
\end{picture}

\caption{Plot of rotationally symmetric solution for $\phi_{\infty}=1$.
Parameters chosen in the figures are: $n=1$, $F=1$, and $\tilde{G}=1/2$
for curved spacetime. The solid and dashed lines correspond to
$|\phi|/v$ and ${\cal J}$.}
\label{fig3}
\end{figure}

\noindent For $r\rightarrow 0$, the behavior of the scalar field
turns out to be
\begin{equation}\label{fr0}
f\approx\left\{
\begin{array}{l}
f_{0}r^{n}\biggl[1+\frac{F}{16}\tilde{G}e^{\tilde{G}}f_{0}^{2\tilde{G}}r^{2}
-\frac{F}{1024}(1+\tilde{G})e^{2\tilde{G}}
\Bigl(16(2+\tilde{G})e^{-\tilde{G}}f_{0}^{2(1-\tilde{G})}-\tilde{G}^{2}F\Bigr)
f_{0}^{4\tilde{G}}r^{4}
+\cdots\,\biggr],\\
~~~~~~~~~~~~~~~~~~~~~~~~~~~~~~~~~~~~~~~~~~~~~~~~~~~~~~~~~~~~~~~~~~
~~~~~~~~~~~~~~~~~~~~~~~~~\mbox{when}\; n=1,\\
f_{0}r^{n}\biggl[1+\frac{F}{16}\tilde{G}e^{\tilde{G}}f_{0}^{2\tilde{G}}r^{2}
-\frac{F^{2}}{512}(1+\tilde{G})e^{2\tilde{G}}f_{0}^{4\tilde{G}}r^{4}
+\cdots\,\biggr],
\,~~~~~~~~~~~~~~~~~~~~~~~\mbox{when}\; n\geq 2.
\end{array}
\right.
\end{equation}
Let us fix an arbitrarily-large number $R_{0}$ for $R$ in
Eq.~(\ref{RRR}). If we choose $f_{0}$ sufficiently small so that
$f_{0}^{\tilde{G}}r^{2} <<1 $ then the terms higher than
second-order in Eq.~(\ref{fr0}) can be neglected for $R<R_{0}$. If
we substitute this approximated solution (\ref{fr0}) into Eq.~(
\ref{fir}) then the energy of the particle ${\cal E}(R_{0})$ in
Eq.~(\ref{fir}) is estimated by
\begin{equation}\label{ptle}
{\cal E}(R_{0})\approx\frac{n^{2}}{2(1-\tilde{G}n)^{2}R_{0}^{2}}
+\frac{F}{2}\tilde{G}n(1-\tilde{G}n)^{\frac{2\tilde{G}n}{1-\tilde{G}n}}
e^{\tilde{G}}
f_{0}^{2\tilde{G}}\Bigl[\frac{1}{R_{0}}-\frac{2}{\tilde{G}^{2}n}\Bigr]
R_{0}^{\frac{2\tilde{G}n}{1-\tilde{G}n}}.
\end{equation}
Let us choose $R_{0}$ and $f_{0}$ such that the square bracket in
the right-hand side of Eq.~(\ref{ptle}) negative and the absolute
value of this second term is larger than the first term.
Furthermore, by taking $f_0$ small enough, additional positive
contribution from the negative higher order terms does not change
the sign of ${\cal E}(R_{0})$.  Then ${\cal E}(R_{0})$ is
negative, which means ${\cal E}(R_{0})<U_{\rm eff}(0)$. Since the
particle energy ${\cal E}$ decreases further as $R>R_{0}$ becomes
large due to friction, the hypothetical particles has a turning
point for  sufficiently small $f_{0}$ and stops finally  at a
local minimum of the effective potential
$u(R=\infty)=\ln[(\tilde{G}+1-\sqrt{2\tilde{G}+1})/\tilde{G}]$
after some oscillation around it. For the large $f_{0}$ and the
sufficiently small $R_{0}$ which makes Taylor expansion in
Eq.~(\ref{fr0}) valid, it is easy to notice that, for $R>R_{0}$,
\begin{equation}\label{urm}
u(R)>2\ln{f_0r^n}+\frac{U'_M}{4}(R_0^2-R^2)+\frac{U'_M}{2}R_0^2\ln{R_0/R}\,,
\end{equation}
where $U'_M=\max_{-\infty<u\le0}(dU_{\rm eff}/du)>0$. The
right-hand side of Eq.~(\ref{urm}) has maximum value at
$R=R_1\equiv
\left[R_0^2+\frac{4n}{U'_M[1-\tilde{G}(n+\tilde{M})]}\right]^{1/2}$
and then $u(R_1)$ satisfies, for a small $R_0$,
\begin{equation}
u(R_1)>\frac{n}{1-\tilde{G}n}
  \left[\ln\frac{4n(1-\tilde{G}n)}{U'_M}-1\right]+2\ln f_0\,.
\end{equation}
Therefore if we choose $f_0$ sufficiently large  as assumed
previously, then $u(R_1)>0$, i.e., the particle went over the
hilltop of the potential at $u=0$. From these results, continuity
now guarantees existence of a vortex solution connecting the
boundary values, $u(0)=-\infty$ and $u(\infty)=0$, for an
appropriate $f_{0}$. This completes the proof and a specific
example $f(=|\phi|/v)$ is shown in Fig.~\ref{fig3}.

Next we discuss that this solution constitutes an asymptotic cone
for $\tilde{G}n<1$ case, however there exists no solution for
$\tilde{G}n>1$ case. When $\tilde{G}n<1$, the radial distance
$\rho$ and the circumference $l$ diverge as $r$ goes to infinity.
 The solutions of Eq. (\ref{fff}) approach their boundary values exponentially
\begin{equation}
f\approx 1-f_{\infty}K_{0}((1-\tilde{G}n)R),
\end{equation}
where $f_{\infty}$ is a constant determined by the proper behavior of the fields
near the origin.
Thus  the asymptotic structure of $\Sigma$ is a cone with deficit
angle $\delta=2\pi\tilde{G}n$ which is flat. With the help of
Eqs.~(\ref{ufu}) -(\ref{RRR}), introduction of a new variable such
as
\begin{equation}
\chi=\ln\psi
\end{equation}
rewrites the equation (\ref{ppp}) as another Newton's equation for
a particle whose position is $\chi$
\begin{equation}\label{eqc}
\frac{d^{2}\chi}{dR^{2}}=-\frac{F}{2}\tilde{G}e^{-\tilde{G}(e^{u}-u-1)}(e^{u}-1)^{2}
-\frac{1}{R}\frac{d\chi}{dR}.
\end{equation}
The external force in the right-hand side of Eq.~(\ref{eqc}) is
always negative so that a particle starts at $\chi=\ln\psi_{0}$
with zero initial velocity and goes to negative infinity as it
decelerates to zero for large $R$. Since the external force term
in the right-hand side of Eq.~(\ref{eqc}) decays exponentially for
large $R$, the leading term of asymptotic solution for $\chi$ is
obtained by solving the linearized equation ~(\ref{eqc}). Suppose
that $\psi\sim r^{-2\alpha}$ ($\alpha\geq0$) for large $r$, these
topological vortices carry quantized magnetic flux ~(\ref{mfr})
\begin{equation}
\Phi=\pm\frac{2\pi n}{e}
\end{equation}
and angular momentum ~(\ref{an2})
\begin{equation}\label{amv}
J=\frac{2\pi\kappa}{e^{2}\tilde{G}}\alpha.
\end{equation}
Note that the spin is arbitrary in curved space
 which is different from the result of flat spacetime, i.e., $J
 \propto n^2$ in flat spacetime.

 Let us consider the case that $\tilde{G}n$ is larger than 1.
Suppose that there exists a solution when $\tilde{G}n>1$.  Then
the radial distance $\rho(r=\infty)$ is finite and the
circumference $l(r=\infty)$ vanishes since the metric $b(r)$
decreases faster than $1/r^{2}$ for large $r$, which means that
the spatial manifold $\Sigma$ is bounded. Since the Euler number
given in Eq.~(\ref{Eul}) must be nonnegative due to the fast
decaying property of the solution, two dimensional sphere $S^{2}$
is  unique candidate. From now on we call the point which
corresponds to $r=0$ ``the south pole" on $S^{2}$ and that which
corresponds to $r=\infty$ ``the north pole" on $S^{2}$. The
configuration of the scalar field $\phi$ at the north pole does
not vanish as $\phi(r=\infty)=ve^{in\theta}$. Therefore the scalar
field $\phi$ is not well-defined at the north pole and there is no
regular solution in this case.

Let us study the case when  $\phi_{\infty}=1$ and $\tilde{G}n=1$.
We introduce $R$ as
\begin{equation}\label{Rln}
R=\ln r,\;\;\;\;\;\;(-\infty<R<\infty)
\end{equation}
which reflects the scale symmetry $(r\rightarrow\lambda r)$ of the
Bogomolnyi equation (\ref{fff}) at this critical value. Then
Eq.(\ref{fff}) takes the same form as Eq.~(\ref{New1}) without
friction term. Similar to the  previous cases, a hypothetical
particle moves under a conservative force only and hence the
Bogomolnyi equation can be integrated to a first-order equation
(\ref{fir})
whose energy ${\cal E}$ of the hypothetical particle is conserved
in this case. The initial particle energy ${\cal E}(R=-\infty)$ is
determined by the initial behavior of $f$ ~(\ref{f0})
\begin{equation}
\left.{\cal E}
   =\frac{1}{2}\left(\frac{du}{dR}\right)^2\right|_{R=-\infty}=2n^2\,.
\end{equation}
Here if we recall that three maxima at $u=0,\;\pm\infty$ are
degenerate, we can easily notice that there is no such particle
motion which satisfies $u=0$ as  $R \to \infty$. Therefore it
completes the nonexistence of soliton solution when
$\phi_{\infty}=1$ and $\tilde{G}n=1$.

\vspace{5mm}

When $\phi_{\infty}\neq 1$, $f$ and $\psi$ are described by the
coupled equations (\ref{fff}) and (\ref{ppp}). As has done in the
case of $\phi_{\infty}=1$, we introduce a new variable
\begin{equation}
\xi=\sqrt{|1-\phi_{\infty}^{2}|}\ln\psi
\end{equation}
with $u$ and $R$ in Eqs.~(\ref{ufu}), (\ref{RRR}) and
(\ref{Rln}). Then the Bogomolnyi equations
(\ref{fff})-(\ref{ppp}) become
\begin{eqnarray}
\frac{d^{2}u}{dR^{2}}&=&-\frac{\partial V_{\rm eff}}{\partial
u}-\frac{1}{R}
\frac{du}{dR},\label{ueq}\\
\frac{d^{2}\xi}{dR^{2}}&=&-\frac{\partial(\mp V_{\rm
eff})}{\partial \xi} -\frac{1}{R}\frac{d\xi}{dR}, \label{chieq}
\end{eqnarray}
where $\mp$ in Eq.~(\ref{chieq}) denotes $-$ for $\phi_{\infty}<1$
cases and $+$ for $\phi_{\infty}>1$ cases. $V_{\rm eff}$ is
\begin{eqnarray}
V_{\rm eff}(u,\xi;\phi_{\infty})&=&U_{\rm eff}(u;\phi_{\infty})W_{\rm eff}(\xi;\phi_{\infty})\nonumber\\
&=&\biggl[-\frac{F}{\tilde{G}}e^{-\tilde{G}(e^{u}-u-1)}
(e^{2u}-2e^{u}+\phi^{2}_{\infty})\biggr]\times\biggl[e^{\tilde{G}
\sqrt{|1-\phi^{2}_{\infty}|}\xi}\biggr].
\end{eqnarray}
Note that Eq.~(\ref{ueq}) contains an impact term at a boundary
value of $R$ which corresponds to $r=0$ if $n\neq0$, and the
friction term in both equations (\ref{ueq})-(\ref{chieq}) vanishes
if $\tilde{G}n=1$, which reflects the scale symmetry
$(r\rightarrow\lambda r)$ of the Bogomolnyi equations
(\ref{fff})-(\ref{ppp}).

 We continue to use the terminology of Newton's equation for the
two-dimensional motion $(u, \xi)$ of a particle when
$\phi_{\infty}>1$ and for one-dimensional motion of two
interacting particles $u$ and $\xi$, when $\phi_{\infty}<1$. The
boundary values of $u$ at $R=\infty$ can be read from shapes of
the  effective potential $U_{\rm eff}$. In order to constitute a
soliton configuration, minimum requirement to the boundary value
of the scalar field, $v\phi_{\infty}$,  at spatial infinity is to
be an extremum of the effective potential $U_{\rm eff}$.
Specifically, from $dU_{\rm eff}/du=0$ and $e^{u(R=\infty)}=
\phi^{2}_{\infty}$, we can easily read that  possible values of
$\phi_{\infty}$ are $0, 1, \displaystyle{\sqrt{1+2/\tilde{G}}} $
and $ \infty$. Since $\phi_{\infty}=1$ case was already analyzed,
let us discuss the other cases. One point on the cosmological
constant should be noted: We recall that
$\phi_{\infty}=0,1,1/\sqrt{\pi Gv^{2}}$ make the potential vanish,
that is, $V(|\phi|=v\phi_{\infty})=0$. If they exist,  regular
solutions satisfying the other  boundary conditions,
$\phi_{\infty}=\sqrt{1+2/\tilde{G}}$ or $\phi_{\infty}=\infty$,
imply self dual solitons in curved spacetime with a nonvanishing
cosmological constant.

\subsection*{\normalsize\bf (b) \underline{$\phi_{\infty}=0$}}
\indent\indent For this case the scalar field approaches the
symmetric local minimum for large $r$ as shown in Fig.~\ref{fig1}
(dashed line). Then the Chern-Simons gauge field is topological
field, i.e., it is no propagating degree, and charged meson is
massive, $m_{\rm meson}=(e^{2}v^{2}/2|\kappa|)$.
  All the  solitonic excitations are the nontopological
solitons which are marginally stable since $E= m_{\rm
meson}Q=(e^{2}v^{2}/2|\kappa|)Q$. We consider those solutions
with nonzero vorticity $(n\ne 0)$ separately from those solutions
with no vorticity $(n=0)$.

\subsection*{\normalsize\bf \hspace{6mm} (b-i) $n=0$}
\indent\indent As mentioned previously, we regard Eqs.~(\ref{ueq})
-(\ref{chieq}) as Newton's equations for two particles of which
the positions are $u(R)$ and $\xi(R)$, respectively. Here
$n\tilde{G}=0$ and then $R$ is in fact equal to $r$ as shown in
Eq.~(\ref{RRR}).  Since the force, $-\partial V_{\rm eff}/\partial
u $, applied to one particle, of which the position is $u$, is
negative definite for $u$ in the region $-\infty<u<0$ and for all
$\xi$, this particle starts out at time zero, $R=0$, at a point
$u_{0}$ $(-\infty<u_{0}<0)$ and approaches to negative infinity
$u=-\infty$,  at infinite time $R=\infty$. Similarly, since the
force $-\partial(-V_{\rm eff})/\partial\xi$ applied to the other
particle, of which the position is, $\xi$ is positive definite for
any $\xi$ and $u$ in the region $-\infty<u<0$, this particle
initially starts out at a point $\xi_{0}$ and finally goes to
positive infinity, $\xi=\infty$.

Near the origin power series solutions give
\begin{eqnarray}
f(r)&\approx&g_{0}\left(1-\frac{F}{16}g_{0}^{2(1+\tilde{G})}(1-g_{0}^{2})
\bigl[2(1+\tilde{G})-\tilde{G}g^{2}_{0}\bigr]e^{\tilde{G}(1-g^{2}_{0})}
\alpha_{0}^{\tilde{G}}r^{2}+\cdots\right),\\
\psi(r)&\approx&\alpha_{0}\Bigl[1+\frac{F}{8}\tilde{G}g_{0}^{2(1+\tilde{G})}
(2-g_{0}^{2})e^{\tilde{G}(1-g^{2}_{0})}
\alpha_{0}^{\tilde{G}}r^{2}+\cdots\Bigr].
\end{eqnarray}
Eqs.~(\ref{fff})-(\ref{ppp}) do not constrain both $g_{0}$
$(0<g_{0}<1)$ and $\alpha_{0}$, however if $g_{0}>1$, both $f(r)$
and $\psi(r)$ will be monotonically-increasing functions of $r$
and hence the boundary condition at $r=\infty$ cannot be met. It
is consistent with the argument given in terms of Newton's
equations for the region of $u_{0}$ $(-\infty<u_{0}<0)$.
Eqs.~(\ref{fff})-(\ref{ppp}) imply that long distance behaviors of
the scalar field $f$ and the off-diagonal component of metric
$\psi$ should  be $f(r)\sim r^{-\varepsilon}$ and $\psi(r)\sim
r^{2\alpha}$ so that Eq.~(\ref{bbr}) gives $b(r)\sim
r^{-2\tilde{G}(\varepsilon-\alpha)}$. Positivity of the energy
~(\ref{BOG}) restricts the values of $\alpha$ and $\varepsilon$ to
satisfy $\varepsilon-\alpha>0$. The precise forms of power series
solutions for large $r$ are
\begin{eqnarray}
f(r)&\approx&g_{\infty}r^{-\varepsilon}\left[1-\frac{F(1+\tilde{G})e^{\tilde{G}}}{
8[\tilde{G}(\varepsilon-\alpha)+\varepsilon-1]^{2}}g_{\infty}^{2(1+\tilde{G})}
\alpha^{\tilde{G}}_{\infty}r^{-2(\tilde{G}(\varepsilon-\alpha)+\varepsilon-1)}+\cdots
\right],\label{gin}\\
\psi(r)&\approx&\alpha_{\infty}r^{2\alpha}\left[1-\frac{F\tilde{G}e^{\tilde{G}}}{
[\tilde{G}(\varepsilon-\alpha)+\varepsilon-1]^{2}}g_{\infty}^{2(1+\tilde{G})}
\alpha^{\tilde{G}}_{\infty}r^{-2(\tilde{G}(\varepsilon-\alpha)+\varepsilon-1)}+\cdots
\right],\label{pin}
\end{eqnarray}
where $g_{\infty}$ and $\alpha_{\infty}$ are the constants
determined by the behavior of the fields near the origin. It is of
our interest to determine the lower bound of
$\tilde{G}(\epsilon-\alpha) +\epsilon-1$ by examining the
correspondence between the short-distance and long-distance
behaviors of the solutions, i.e., $(g_{0}, \alpha_{0})$ and
$(\epsilon,\alpha)$. Since $f$ becomes small near the origin for a
sufficiently-small $g_{0}$ and an arbitrary $\alpha_{0}$, we can
expand the right-hand side of Eqs.~(\ref{fff})-(\ref{ppp}) for the
leading order of $f$. If we define $g$ as
$g(r)=Fe^{\tilde{G}}(1+2\tilde{G})f^{2(1+\tilde{G})}\psi^{\tilde{G}}$,
then the approximate equation for $g$ is the Liouville equation.
An exact solution  satisfying the boundary condition at the origin
is
\begin{equation}\label{grr}
g(r)=\frac{g_{0}^{2(1+\tilde{G})}\alpha_{0}^{\tilde{G}}}{(1+\frac{1}{8}
g_{0}^{2(1+\tilde{G})}\alpha_{0}^{\tilde{G}}r^{2})^{2}},
\end{equation}
which is small enough for all $r$ and approaches to the exact
solution of Eqs.~(\ref{fff})-(\ref{ppp}) as $g_{0}\rightarrow0$.
Comparing the solution ~(\ref{grr}) with those in
Eqs.~(\ref{gin})-(\ref{pin}), we obtain a lower bound for
$\epsilon$ and $\alpha$, i.e.,
$\tilde{G}(\epsilon-\alpha)+\epsilon-1\geq1$, and it is consistent
with  regularity of the power series solutions
~(\ref{gin})-(\ref{pin}). As $g_{0}\rightarrow0$, the magnetic
flux becomes $\Phi=\pm (2-\varepsilon)/2eGv^{2}$ and it implies a
constraint for $\varepsilon$, i.e., $\varepsilon\le 2$.

Furthermore, another category in examining the solutions is to
know that  the area ${\cal A}$ of spatial manifold $\Sigma$ is
finite or infinite. When $\tilde{G}(\varepsilon-\alpha)<1$, the
area ${\cal A}$, the radial distance at infinity $\rho(r=\infty)$,
and the circumference at infinity $l(r=\infty)$ are infinite.
Then the asymptotic structure of $\Sigma$ is also a cone with
deficit angle $\delta=8\pi^{2}Gv^{2}(\varepsilon-\alpha)$. Though
the global structure of the cone is the same as that of vortices
of $\phi_{\infty}=1$ case, these manifolds have long-range tails
for large $r$ as shown in Fig.~\ref{fig4}-(a) and the shape of the
angular momentum density (see Fig.~\ref{fig4}-(b)) shows that it
is accumulated near the origin. Since the circumference at
infinity $l(r=\infty)$ is finite  when
$\tilde{G}(\varepsilon-\alpha)=1$, the asymptotic structure of
$\Sigma$ is a cylinder.
When $\tilde{G}(\varepsilon-\alpha)>1$, the space $\Sigma$ is
bounded  manifold and then a two sphere is  unique candidate. The
Euler number fixes $\tilde{G}(\varepsilon-\alpha)=2$ that is
$b(r)\sim r^{-4}$. Then the regularity at the north pole
$(r=\infty)$ requires that the scalar field behaves as $f\sim
s^{n}$ for $s\sim 0$, where $s$ is radial coordinate in the new
coordinate system whose origin is at the north pole. However,
since the scalar field $\phi$ does not vanish at the south pole
which corresponds to the point $r=0$ or equivalently $s=\infty$,
it is not well-defined. Specifically, $\phi=|\phi|
e^{in\theta}=g_{0}e^{in\theta}$ at the north pole. Therefore there
is no regular solution when $\tilde{G}(\varepsilon-\alpha)>1$.
\begin{figure}
\vspace{-8mm}

\setlength{\unitlength}{0.1bp}
\begin{picture}(3600,2807)(0,0)
\put(2000,0){\makebox(0,0){\Large $\tilde{r}$}}
\put(-100,1553){%
\makebox(0,0)[b]{\shortstack{\Large $|\phi|/v$}}%
}
\put(3550,300){\makebox(0,0){40}}
\put(2775,300){\makebox(0,0){30}}
\put(2000,300){\makebox(0,0){20}}
\put(1225,300){\makebox(0,0){10}}
\put(450,300){\makebox(0,0){0}}
\put(400,2707){\makebox(0,0)[r]{0.8}}
\put(400,2130){\makebox(0,0)[r]{0.6}}
\put(400,1554){\makebox(0,0)[r]{0.4}}
\put(400,977){\makebox(0,0)[r]{0.2}}
\put(400,400){\makebox(0,0)[r]{0}}
\end{picture}

\vspace{-7mm}
\begin{center}{(a)}
\end{center}

\setlength{\unitlength}{0.1bp}
\begin{picture}(3600,2807)(0,0)
\put(2025,0){\makebox(0,0){\Large $\tilde{r}$}}
\put(0,1553){%
\makebox(0,0)[b]{\shortstack{\Large ${\cal J}$}}%
}
\put(3550,300){\makebox(0,0){40}}
\put(2788,300){\makebox(0,0){30}}
\put(2025,300){\makebox(0,0){20}}
\put(1263,300){\makebox(0,0){10}}
\put(500,300){\makebox(0,0){0}}
\put(450,2707){\makebox(0,0)[r]{0.25}}
\put(450,2246){\makebox(0,0)[r]{0.2}}
\put(450,1784){\makebox(0,0)[r]{0.15}}
\put(450,1323){\makebox(0,0)[r]{0.1}}
\put(450,861){\makebox(0,0)[r]{0.05}}
\put(450,400){\makebox(0,0)[r]{0}}
\end{picture}

\vspace{-7mm}
\begin{center}{(b)}
\end{center}
\vspace{-4mm} \caption{Plot of rotationally symmetric
nontopological solitons for $\phi_{\infty}=0$ and $n=0$.
Parameters chosen in the figures are $\varepsilon-\alpha=1$,
$F=1$, $\tilde{G}=1/2$ for curved spacetime: (a) $|\phi|/v$ vs
$\tilde{r}$ (b) ${\cal J}$ vs $\tilde{r}$ . The solid and dashed
lines correspond to $f(0)=0.8$ and $0.3$, respectively.}
\label{fig4}
\end{figure}

Since the solutions with $|\phi|(\infty)=0$ are nontopological
solitons, they are characterized by the U(1) charge (or
equivalently the magnetic flux)
$Q(=-\kappa\Phi)=\mp(2\pi\kappa/e)\Bigl[\varepsilon+(e^{2}\tilde{G}/2\pi\kappa)
J\Bigr]$ and the angular momentum
$J=-(2\pi\kappa/e^{2}\tilde{G})\alpha$, which need not be
quantized. Note that the sign of angular momentum for these
nontopological solitons is opposite to that of vortices as given
in Eq.~(\ref{amv}) as in the case of  flat spacetime. In case of
the  cylinder the magnetic flux has a fixed value
$\Phi=2\pi\kappa/e\tilde{G}$ for a given set of parameters.

\subsection*{\normalsize\bf \hspace{6mm} (b-ii) $n\neq 0$}
\indent\indent For $n\neq0$ solutions, the motion of a
hypothetical particle whose position is depicted by $\xi$
coordinate resembles that of the $n=0$ solution since it follows
the same equation of motion (\ref{chieq}) and the starting point
is also an arbitrary constant $\xi_{0}$. On the other hand, the
motion of the other hypothetical particle whose position is $u$
coordinate is different from $n=0$ solution due to the vorticity.
Specifically, the particle should start out at negative infinity,
$u=-\infty$,  with an initial velocity given by Eq.~(\ref{f0})
which is induced by receiving an initial impact at time
corresponding to $r=0$, and then it turns at an appropriate
position $u_{\rm max}$ at a certain time, and finally returns to
$u=-\infty$ at time $R(r)|_{r=0}$. We now argue that there exist
such nontopological vortex solutions by showing the existence of
turning point $u_{\rm max}$ between $-\infty$ and 0.

For small $r$,  power series solutions are
\begin{eqnarray}
f(r)&\approx&h_{0}r^{n}\Bigl[1-\frac{F}{8(n+1)^{2}}(1+\tilde{G})e^{\tilde{G}}
h_{0}^{2(1+\tilde{G})}\beta_{0}^{\tilde{G}}r^{2(n+1)}+\cdots\Bigr],
\label{pff}\\
\psi(r)&\approx&\beta_{0}\Bigl[1+F\tilde{G}e^{\tilde{G}}h_{0}^{2(1+\tilde{G})}
\beta_{0}^{\tilde{G}}r^{2(n+1)}+\cdots\Bigr].
\label{pps}
\end{eqnarray}
Note that the sign of second-order terms is negative for $f$ and
positive for $\psi$, which looks consistent with the above
description in terms of Newtonian mechanics. Let us assume an
arbitrarily-large number $R_{0}$ for $R$ both in Eq.~(\ref{pff})
and Eq.~(\ref{pps}). If we choose $h_{0}$ sufficiently small for a
given $\beta_{0}$, then the terms higher than second-order both in
Eq.~(\ref{pff}) and Eq.~(\ref{pps}) can be neglected for $R\leq
R_{0}$.

The leading terms of $f$ and $\psi$ for large $r$ are assumed to
be $f(r)\sim r^{-\varepsilon}$ $(\epsilon>0)$ and $\psi(r)\sim
r^{2\alpha}$ $(\alpha\geq0)$ so that we obtain a condition
$n+\varepsilon-\alpha>0$ from positivity of the energy. Long
distance behavior of $f$ and $\psi$ is
\begin{eqnarray}
f(r)&\approx&h_{\infty}r^{-\varepsilon}\left[1-\frac{F(1+\tilde{G})e^{\tilde{G}}}{
8[\tilde{G}(n+\varepsilon-\alpha)+\varepsilon-1]^{2}}h_{\infty}^{2(1+\tilde{G})}
\psi^{\tilde{G}}_{\infty}r^{-2(\tilde{G}(n+\varepsilon-\alpha)+\varepsilon-1)}+\cdots
\right],\\
\psi(r)&\approx&\psi_{\infty}r^{2\alpha}\left[1-\frac{F\tilde{G}e^{\tilde{G}}}{
[\tilde{G}(n+\varepsilon-\alpha)+\varepsilon-1]^{2}}h_{\infty}^{2(1+\tilde{G})}
\beta^{\tilde{G}}_{\infty}r^{-2(\tilde{G}(n+\varepsilon-\alpha)+\varepsilon-1)}+\cdots
\right],
\end{eqnarray}
where $h_{\infty}$ and $\beta_{\infty}$ are the constants fixed by
requiring the proper behavior of the fields near the origin.
Through an approximation to the Liouville equation as was done in
$n=0$ case,  we obtain a condition
$\tilde{G}(n+\varepsilon-\alpha)+\epsilon-1 \geq n+1$ for each $n$
and the equality holds when $h_{0}\rightarrow0$. In this limit the
magnetic flux become $\Phi=\pm (n-\varepsilon+1)/2eGv^{2}$ which
implies the upper bound of $\varepsilon$, i.e., $\varepsilon\leq
n+2$.

 Analysis of the asymptotic structure of spatial manifold
$\Sigma$ is almost the same as $n=0$ case, if we replace
$\varepsilon-\alpha$ to $n+\varepsilon-\alpha$. The $n\neq 0$
solitonic configurations support asymptotically a cone with
deficit angle $\delta=8\pi^{2}Gv^{2}(n+\varepsilon-\alpha)$ when
$\tilde{G}(n+ \varepsilon-\alpha)<1$ and a cylinder when
$\tilde{G}(n+\varepsilon-\alpha)=1$. On these open spaces, the
nontopological vortices are  hybrids of the vortices at
short-distance region and the nontopological solitons at
long-distance region as shown in Fig. 5-(a)(b).
\begin{figure}
\vspace{-8mm}

\setlength{\unitlength}{0.1bp}
\begin{picture}(3600,2807)(0,0)
\put(2000,150){\makebox(0,0){\Large $\tilde{r}$}}
\put(100,1553){%
\makebox(0,0)[b]{\shortstack{\Large $|\phi|/v$}}%
}
\put(3550,300){\makebox(0,0){800}}
\put(2775,300){\makebox(0,0){600}}
\put(2000,300){\makebox(0,0){400}}
\put(1225,300){\makebox(0,0){200}}
\put(450,300){\makebox(0,0){0}}
\put(400,2707){\makebox(0,0)[r]{0.6}}
\put(400,1938){\makebox(0,0)[r]{0.4}}
\put(400,1169){\makebox(0,0)[r]{0.2}}
\put(400,400){\makebox(0,0)[r]{0}}
\end{picture}

\vspace{-7mm}
\begin{center}{(a)}
\end{center}

\setlength{\unitlength}{0.1bp}
\begin{picture}(3600,2807)(0,0)
\put(2025,150){\makebox(0,0){\Large $\tilde{r}$}}
\put(100,1553){%
\makebox(0,0)[b]{\shortstack{\Large ${\cal J}$}}%
}
\put(3550,300){\makebox(0,0){600}}
\put(3042,300){\makebox(0,0){500}}
\put(2533,300){\makebox(0,0){400}}
\put(2025,300){\makebox(0,0){300}}
\put(1517,300){\makebox(0,0){200}}
\put(1008,300){\makebox(0,0){100}}
\put(500,300){\makebox(0,0){0}}
\put(450,2707){\makebox(0,0)[r]{0.15}}
\put(450,2246){\makebox(0,0)[r]{0.12}}
\put(450,1784){\makebox(0,0)[r]{0.09}}
\put(450,1323){\makebox(0,0)[r]{0.06}}
\put(450,861){\makebox(0,0)[r]{0.03}}
\put(450,400){\makebox(0,0)[r]{0}}
\end{picture}

\vspace{-7mm}
\begin{center}{(b)}
\end{center}
\vspace{-4mm} \caption{Plot of rotationally symmetric solutions
for $\phi_{\infty}=0$ and $n=1$. Parameters chosen in the figures
are $\varepsilon-\alpha=1$, $F=1$, $\tilde{G}=1/2$ for curved
spacetime. (a) $|\phi|/v$ vs $\tilde{r}$ (b)  ${\cal J}$ vs
$\tilde{r}$. The solid and dashed lines correspond to $(h_\infty,
\psi_\infty)=(0.02,0.2)$ and (0.0017,0.2), respectively.}
\label{fig5}
\end{figure}

When $\tilde{G}(n+\varepsilon-\alpha)>1$, the space $\Sigma$ is
bounded and then it should form a two dimensional sphere. Then the
Euler invariant ~(\ref{Eul}) should also be equal to that of
 a smooth $S^{2}$ such as
\begin{equation}
\tilde{G}(n+\varepsilon-\alpha)=2.
\end{equation}
Now we show that both $\varepsilon$ and $\alpha$ can not be
arbitrary by imposing the regularity condition. From the behavior
of the metric $b(r)$ at large $r$, the radial distance $\rho(r)$
behaves near the north pole as $\rho-\rho(\infty)\sim
r^{-\tilde{G}(n+\varepsilon-\alpha)+1}$. On the other hand, if we
choose a new radial coordinate $t$ around the north pole as
$\rho-\rho(\infty)\sim t$, then the regularity demands that the
scalar field behaves as $f\sim t^{n}$ for small $t$. Comparing
these with Eq.~(\ref{f0}), we get $\varepsilon=n$ and
$\displaystyle{ \alpha=2\Bigl(n-\frac{1}{\tilde{G}}\Bigr)}$.
Moreover, since $\psi$ expressed in $t$ coordinate should also be
a positive constant at the north pole, $\alpha$ (equivalently the
angular momentum $J$) should vanish and thereby the solution can
exist only when $n=1/\tilde{G}$. For this sphere case
 with $\tilde{G}n=1$ the Bogomolnyi equations in Eq.~(\ref{ueq})-(\ref{chieq})
 have no friction term. Therefore, one of those can be
integrated to a first-order equation
\begin{equation}
\frac{1}{2}\biggl(\frac{du}{dR}\biggr)^{2}-\frac{1}{2}\biggl(\frac{d\xi}{dR}
\biggr)^{2}+V_{\rm eff}(u,\xi )=2n^{2},
\end{equation}
where the integration constant in the right-hand side is
determined by considering the behaviors of $u$ and $\xi$ near
$r=0$ in Eqs.~(\ref{pff})-(\ref{pps}). This sphere is symmetric
under the inversion between the south pole and the north pole  so
that it can also be interpreted as a configuration that two
vortices with vorticity $n$ lie both at the south pole and at the
north pole.  The total magnetic flux is twice of the flux of a
hemisphere,
$\displaystyle{\Phi=\pm\frac{4\pi}{e}n=\pm\frac{1}{Gev^{2}}}$.

\vspace{5mm}

\subsection*{\normalsize\bf (c) \underline{$\phi_{\infty}=\displaystyle{
\sqrt{1+\frac{2}{\tilde{G}}}}$}} \indent\indent Suppose that for
each $n$ there exists a solution to satisfy this boundary
condition. Since the shape of effective potential $U_{\rm eff}$
given by the dotted line in Fig.~\ref{fig2}-(a)  may guarantee the
existence of a solution, the finite energy condition in Eq.~
(\ref{eee}) restricts $\tilde{G}(n+\alpha(1-\phi^{2}_{\infty}))>1$
and it implies the volume of two dimensional space $\Sigma$ should
be finite.  Two sphere is the unique candidate as explained in
$\phi_{\infty}=0$ case. Since $\phi_{\infty}\neq0$ contradicts to
the condition for the regularity on the north pole of $S^{2}$,
there does not exist such regular solution.

\vspace{5mm}

\subsection*{\normalsize\bf (d) \underline{$\phi_{\infty}=\infty$}}
\indent\indent If we look at the scalar potential unbounded below
as given in Fig.~\ref{fig1} and an expression of the  energy in
Eq.~(\ref{eee}), we notice that there is no reason to exclude
$\phi_{\infty}=\infty$ solution as a candidate of a finite-energy
solution. When $n=0$, $u=\infty$ is the solution which describes
the potential minimum at negative infinity. However the energy of
it is of course negative infinite  so that it is the solution out
of our scope. When $n\neq0$, the conservative force by the
effective potential, $dU_{\rm eff}/du$, is singular at $u=0$ and
thereby there is no regular solution to connect both boundaries
$u=-\infty$ and $u=\infty$.

\setcounter{equation}{0}
\section{Behaviors of Multi-soliton Solutions}
In the previous section we restricted our interest to self-dual
solitons with rotational symmetry.  We study the behaviors of
arbitrary multi-soliton configurations without rotational symmetry
in what follows.

First, we show the fast decaying property of topological vortices
on asymptotic region irrespective of both parameters of the theory
and positions of vortex centers. Then nonexistence of the
topological multi-vortices are  proved when $\tilde{G}n>1$.

We prove the fast decaying property of solutions of
Eq.~(\ref{Bogm}), which is not necessarily radial. \vskip 0.2 true
cm
 \noindent
 \bf Theorem 1:\rm \quad
  Assume that a  solution  $f^2$ for Eq.~(\ref{Bogm}) satisfies the
following conditions: ~$0\le f^2 \le 1$,  $
\mathop{\lim}  \limits_{z\to \infty}  f^2  = \phi_{\infty}=1,~~
 0 <{\tilde{G}}n <1,   ~~ $ and $e^{h+\overline{h}}=c_0 $ is a positive
constant. Then $f^2$ converges to $1$ faster than any polynomial
rate as $|z|$ increase in the following sense: $f^2$ satisfies
$|f^2(z)-1| < {1 \over |z|^k}$ as $|z| \to \infty$ for any
positive $k$. \vskip 0.3 true cm

Next we show the following  nonexistence of multi-vortices. \vskip
0.2 true cm
 \noindent
 \bf Theorem 2:\rm \quad Assume that $0\le f^2 \le 1$, $ {\tilde{G}}n >1 , ~~
\lim_{z\to \infty} f^2=\phi_{\infty}=1 $,  and
$e^{h+\overline{h}}=c_0 $ is a positive constant in Eq.
(\ref{Bogm}). Then  there is no solution $f^2$ of
Eq.~(\ref{Bogm}) with $\int dz |f^{2}-1|^{2}<\infty$.

\vskip 0.3 true cm Detailed proofs of Theorem 1 and  2 are
presented in Appendix.

\setcounter{equation}{0}
\section{Conclusion}
In this paper we have precisely described the self-dual
Chern-Simons solitons in curved spacetime. Bogomolnyi type bound
for the original and the dual-transformed theory has been derived
under a one-parameter family of $\phi^{8}$ scalar potential. In
the context of the duality transformation in continuum, the role
of Higgs and the topological sector of scalar phase have been
clearly demonstrated by showing how they support the
nonperturbative excitations, specifically the soliton spectra,
and affect to their mutual interactions through  introduction of a
Jacobian in path integral measures. Then we analyzed the
Bogomolnyi equations under a rotationally symmetric  ansatz and
obtained all possible soliton solutions. Although the scalar
potential has an additional $\phi^{8}$ term with negative
coefficient in comparison with that in flat spacetime and changes
its form along the boundary value of  the scalar field, this
system supports regular soliton configurations with finite,
positive energy whose boundary values are the same as those in
flat spacetime, i.e., $|\phi|(\infty)=v$ or $|\phi|(\infty)=0$.
In the former case there are topological vortices which carry a
quantized magnetic flux (or a U(1) charge) and constitute an
asymptotic cone  and cylinder  as the underlying spatial
manifold. The latter case contains the nontopological solitons
and vortices which have a continuous variable charge and whose
underlying manifolds comprise an asymptotic cone, an asymptotic
cylinder, and a two sphere as the value of total magnetic flux is
less than or equal to a critical value. All of them are spinning
objects whose values can be arbitrary in contrary to the vortices
in flat spacetime.

While the existence and the properties of rotationally-symmetric
self-dual solitons have been rigorously examined, the stability
of those solutions is an open question. Perhaps the unboundedness
of scalar potential can be understood by a supergravity version
of the model as the supersymmetry did in flat
spacetime~\cite{LLW}, but the stability of nontopological objects
outside the Bogomolnyi limit both in flat and curved spacetime
may await investigation using numerical computation. In relation
with the cosmological constant, asymptotic region of every open
space formed by the self-dual solitons has zero cosmological
constant though there exist several local vacua with nonvanishing
cosmological constant. It may imply a connection between the
self-dual Chern-Simons solitons and supergravity theory
~\cite{BBS}. This seems also consistent with unattainability of a Bogomolnyi
type bound in a gauged system with nonvanishing cosmological
constant~\cite{KK2}.
 For multi-soliton solutions, which are not necessary rotationally
 symmetric, fast decaying property is proved and nonexistence of
 finite energy solutions are presented when $\tilde{G}n >1$.
Though some aspects of solitons in dual-transformed theory were
briefly commented, the quantum field theoretic issues such as the
phase transition induced by Chern-Simons solitons need further
study.

\renewcommand{\theequation}{A.\arabic{equation}}
\setcounter{equation}{0}
\section{Appendix}
In Appendix, detailed mathematical proofs of Theorem 1 and  2 are
provided.
  We denote $dz=dx_1 dx_2$, $\Delta=\tilde{\partial}^2,$
$a={{1+{\tilde{G}}-\sqrt{1+2{\tilde{G}}}}\over {\tilde{G}}}$, $d=
{{1+{\tilde{G}}+\sqrt{1+2{\tilde{G}}}}\over {\tilde{G}}}$ and \bea
H(z,T)=H_1(z,T)+H_2(z,T) ,\nonumber \eea \noindent where \bea
H_1(z,T)=\triangle~ \ln~T \nonumber \eea \noindent and \bea
H_2(z,T)= {{\tilde{G} c_0}\over 2}{{T^{\tilde{G}}
e^{{\tilde{G}}(1-T)}}\over { (\prod_{p=1}^{p=n}
|z-z_p|^2)^{\tilde{G}} }} (T-1) (T-a) (T-d). \nonumber \eea When
$e^{h+\overline{h}}=c_0$ and $ \phi_{\infty}=1$, Eq.~(\ref{Bogm})
is equivalent to $H(z, T)=0$. Let $ H_1^2(R^2) $ be the Sobolev
space, which is the completion of $C_c^{\infty}(R^2)$ with respect
to the norm $||w||=\big( \int_{R^2} |\nabla w|^2 +w^2 \ dz
)^{1\over 2}$.
\vskip 0.3 true cm \noindent  \it Proof of Theorem 1. \vskip 0.2
true cm \rm
 Take $u(z)=1 -c/|z|^k$. We
will show that $u \le f^2$ on the outside of a large ball. \par
\noindent \it Step 1. \rm \quad $H(z,u)\ge 0$ on the outside of a
large ball.

 Take a fixed constant $b$ so that
$a <b<1$ and \bea {{\partial} \over {\partial T}} H_2(z,T) \le 0 ,
\label {eq10} \eea when $b\le T \le 1$ and $|z|$ is large. Since
$f^2$ goes to $1$ at infinity, we can take large $r_0$ so that
$f^2(z) \ge b$ on $|z|\ge r_0$ and we take $c$ so that
$u(r_0)=1-c/r_0^k=b$. Note that \bea \Delta~ \ln ~ u&=&-{{c k^2
|z|^{k-2}}\over (|z|^k-c)^2} \nonumber \\&=& -{{ck^2}\over
{|z|^{k+2} u^2}} . \nonumber \eea \noindent We calculate $H(z,u)$
on $|z|\ge r_0$. Since $|u-a|\ge |b-a|$ and $|u-d| \ge |d-1|$,
\bea H_2(z,u)&=& {c_0{\tilde{G}}\over 2}{{u^{\tilde{G}} e^{(1-u)
{\tilde{G}}}}\over { (\prod_{p=1}^{p=n} |z-z_p|^2)^{\tilde{G}} }}
(u-1) (u-a) (u-d)\nonumber \\
&\ge& {c_0{\tilde{G}} c \over {2 |z|^k}}   {{u^{\tilde{G}}
 \over { (\prod_{p=1}^{p=n} |z-z_p|^2)^{\tilde{G}} }}}
(b-a) (d-1) . \nonumber \eea Let us denote
$\alpha={\tilde{G}}(b-a)(d-1)c_0/2$. Since  $( \prod_{p=1}^{p=n}
|z-z_p|^2)^{\tilde{G}}  \le c'|z|^{2{\tilde{G}}n}$ for some
constant $c'$ and  $|z|\ge r_0$, we estimates $H(z,u)$
 \bea H(z,u)
&\ge& {c \over |z|^{k+2}} \left(\alpha u^{\tilde{G}} {|z|^2 \over
{
(\prod_{p=1}^{p=n} |z-z_p|^2)^{\tilde{G}} }} -k^2 /u^2 \right )\nonumber \\
&\ge& {c \over |z|^{k+2}} \left(\alpha  b^{\tilde{G}} {|z|^2 \over
{
(\prod_{p=1}^{p=n} |z-z_p|^2)^{\tilde{G}} }} -k^2 /b^2 \right ) \nonumber \\
&\ge& {c \over |z|^{k+2}} \left(\alpha c' b^{\tilde{G}}
|z|^{2(1-{\tilde{G}}n)} -k^2 /b^2 \right ) . \label{eq111}  \eea
\noindent Eq.~(\ref{eq111}) is nonnegative when $|z|\ge r_0$ by
taking $b$ which is sufficiently to $1$. \vskip 0.3 true cm

\it Step 2. \quad \rm $f^2(z) \ge u(z)$  on $ D\equiv R^2-B(r_0)$,
where $r_0$ is defined in step 1. \vskip 0.3 true cm

 Assume that $f^2=e^w$ and $u=e^m$
on $D$, where $w$ and $m$ are smooth functions on $R^2$.
 Note that $H(z,f^2)=H(z, e^w)=0$, $H(z,u)=H(z,e^m)\ge 0$ on
 $D$, and $f^2(z)\ge u(z)=b$ on $\partial D$. From Eq.~(\ref{eq111}),
 we have
\bea 0 &\ge& H(z, f^2)-H(z,u) \nonumber \\&=&H(z,e^w)-H(z,e^m)\\
\nonumber &=& \Delta (w-m) +H_2(z,f^2)-H_2(z,u) .  \nonumber \eea
\noindent
{}From Eq.~(\ref{eq10}),
\bea H_2(z,T_1)-H_2(z,T_2)=-\lambda(z) (T_1-T_2) \nonumber  \eea
for some nonnegative function $\lambda(z)$ for $b\le T_1, T_2 \le
1 $. Therefore, \bea \Delta (w-m) \le  -H_2(z,f^2)+H_2(z,u) =
\lambda(z) (f^2-u) . \label {m1}\eea \noindent Assume that $f^2-u
< 0$ on a domain $D' \subset D$, then $\Delta (w-m) \le 0$ on
$D'$. By the Maximum Principle $w-m$ cannot have minimum inside
of $D'$. Since $w=m$ on $\partial D'$ and $w< m$ on $D'$, $D'$
must be the empty set. We conclude that $f^2 \ge u$ on $D$. We
proved Theorem 1.

\vskip 0.3 true cm
 Assume that there exists $w\in H_1^2(R^2)$ so that the solution
$f^2=e^w \le 1$ for Eq.~(2.22) on $D$. We claim that $f^2$
converges to $1$ as $|z|\to \infty$.

\noindent From Eq.~(\ref{Bogm}), \bea \Delta w=-H_2(z,e^w) .
\nonumber \eea  For a point $q \in D $,   $ |w(q)|\le C \left(
\int _{B(q,1)} |w|^2 +H_2(z,e^w)^2 dz \right) $ by  the  standard
elliptic estimates for $w$ on $R^2$ (see \cite{Gil}). Since $0\le
T=f^2=e^w\le 1$, $ \int _{B(q,1)} H_2(z,e^w)^2 dz  $ goes to $0$
as $|q|\to \infty$.
 Since $w\in
H_1^2(R^2)$, $|w(q)| \to 0$ as $|q| \to \infty$. Therefore,
$f^2(z) \to 1$ as $|z| \to \infty$. From the above we proved the
following Theorem. \vskip 0.3 true cm \noindent

\bf Theorem A:\rm \quad Assume that $ {\tilde{G}}n <1 , ~~
\phi_{\infty}=1 $, $e^{h+\overline{h}}=c_0 $ is a positive
constant, and there exists $w\in H_1^2(R^2)$ so that a solution
$f^2=e^w \le 1$ for Eq.~(\ref{Bogm}) on $D=R^2-B(R_1)$, where $D$
is the outside of some large ball of radius $R_1$. Then $f^2$
converges to $1$ faster than any polynomial rate as $|z|$
increases in the following sense: $f^2$ satisfies $|f^2(z)-1| < {1
\over |z|^k}$ as $|z| \to \infty$ for any positive $k$.

\vskip 0.3 true cm \noindent \it Proof of Theorem 2. \rm \vskip
0.2 true cm

We use the same notations as in the Proof of Theorem 1. Take
$u(z)$, $b$, and $r_0$ as in the Theorem 1. Note that $f^2(z) \ge
b$ when $|z|\ge r_0$.  Choose $b'$ so that $b<b'<1$. Take
$r_1>r_0$ so that $f^2(z)\le b'$ on $|z|=r_1$. Take $c$ so that
$u(r_1)=1-c/r_1^k=b'$. Since $|u-a|<|1-a|$ and $|u-d|<|d-b|$, \bea
H_2(z,u)&=& {c_0{\tilde{G}}\over 2}{{u^{\tilde{G}} e^{(1-u)
{\tilde{G}}}}\over { (\prod_{p=1}^{p=n} |z-z_p|^2)^{\tilde{G}} }}
(u-1) (u-a) (u-d)
\nonumber \\
&\le& {c_0{\tilde{G}} c \over {2 |z|^k}}   {{u^{\tilde{G}}
 \over { (\prod_{p=1}^{p=n} |z-z_p|^2)^{\tilde{G}} }}}
(1-a) (d-b).  \nonumber \eea \noindent
 Let us denote $\alpha'={\tilde{G}}(1-a)(d-b)c_0/2$. Since
 $(\prod_{p=1}^{p=n} |z-z_p|^2)^{\tilde{G}}  \ge
\bc |z|^{2{\tilde{G}}n}$ for some constant $\bc$ and  $|z|\ge
r_1$, we estimate $H(z,u)$,
 \bea H(z,u)
&\le& {c \over |z|^{k+2}} \left[\alpha' u^{\tilde{G}} {|z|^2 \over
{
(\prod_{p=1}^{p=n} |z-z_p|^2)^{\tilde{G}} }} -k^2 /u^2 \right ]\nonumber \\
&\le& {c \over |z|^{k+2}} \left[\alpha'   {|z|^2 \over {
(\prod_{p=1}^{p=n} |z-z_p|^2)^{\tilde{G}} }} -k^2  \right ] \nonumber \\
&\le& {c \over |z|^{k+2}} \left(\alpha' \bc
|z|^{2(1-{\tilde{G}}n)} -k^2  \right ) . \label{eq11}  \eea
\noindent Since $\tilde{G}n >1$, Eq.~(\ref{eq11}) is nonpositive
when $|z|\ge r_1 $ by taking $b'$ which is sufficiently close to
$1$. \vskip 0.3 true cm

Next we show that $f^2(z) \le u(z)$  on $ D=R^2-B(r_1)$, where
$r_1$ is defined in above. \vskip 0.3 true cm \noindent
 Assume that $f^2=e^w$ and $u=e^m$
on $D$, where $w$ and $m$ are smooth functions on $R^2$.
 Note that $H(z,f^2)=H(z, e^w)=0$, $H(z,u)=H(z,e^m)\le 0$ on
 $D$, and $f^2(z)\le u(z)=b'$ on $\partial D$.
{}From above, we have
\bea 0 &\le& H(z, f^2)-H(z,u) \nonumber \\&=&H(z,e^w)-H(z,e^m)\\
\nonumber &=& \Delta (w-m) +H_2(z,f^2)-H_2(z,u).  \nonumber \eea
\noindent
{}From Eq.~(\ref{eq10}),
\bea H_2(z,T_1)-H_2(z,T_2)=-\lambda(z) (T_1-T_2) \nonumber  \eea
for some nonnegative function $\lambda(z)$ for $b\le T_1, T_2 \le
1 $. Therefore, \bea \Delta (w-m) \ge  -H_2(z,f^2)+H_2(z,u) =
\lambda(z) (f^2-u) \label {mm1}\eea \noindent Assume that $f^2-u >
0$ on a domain $D' \subset D$, then $\Delta (w-m) \ge 0$ on $D'$.
By the Maximum Principle $w-m$ can not have maximum inside of
$D'$. Since $w=m$ on $\partial D'$ and $w> m$ on $D'$, $D'$ must
be the empty set. We conclude that $f^2 \le u=1-c/|z|^k$ on $D$.

For $w\in H_1^2(R^2)$, $\int_{R^2} |e^w-1|^2 dz \le c \exp (c_1
\int_R^2 |\nabla w|^2+w^2 dx)$ for some constant $c$ and $c_1$
(see \cite{Wan}). Therefore, $|e^w-1|^2=|f^2-1|>c^2/|z|^{2k} $
cannot hold for $k\le 2$ as $|z|\to \infty$. Thus we proved
Theorem 2.

\vskip 0.3 true cm
 \it Remark\rm \quad When $\phi_\infty=1$,
$\Delta ~\ln~\psi \le 0$ by Eq.~(2.23). Since there is no lower
bounded super harmonic function on $R^2$, there is no constant
$c$ such that $\psi>c^2>0$ and $\psi$ has no local minimum.
\vskip 0.3 true cm Next we study the behavior of $f^2$ when
$\phi_\infty=0$ and $0 \le f^2 \le 1$. When $\phi_\infty=0$,
$\Delta ~\ln~\psi \ge 0$ by Eq.~(\ref{eqK}). Since there is no
upper bounded subharmonic function on $R^2$, there is no constant
$c$ such that $\psi < c^2$ and $\psi$ has no local maximum. From
Eq.~(\ref{eqK}), we obtain a decaying condition for $f^2$ in the
following.

\vskip 0.3 true cm \bf Claim \rm Let $f^2$ be a solution for
Eqs.~(\ref{Bogm})-(\ref{eqK}) when $e^{h+\overline{h}}=c_0 $ is a
positive constant and $\phi_\infty=0$.  There is no positive
constant $c$ such that $f^2 \ge c |z|^{ {2(n\tilde{G}-1)}\over
{1+ \tilde{G}} }$ for large $|z|$.

\it Proof of Claim: \rm \quad Take $\psi^{ \tilde{G} } =e^h $ in
Eq.~(\ref{eqK}). For large $|z|$, Eq.~(\ref{eqK}) turns into the
following equation
 \bea \Delta h&=& {{\tilde{G}^2 c_0}\over
2}{{f^{2 \tilde{G}} e^{{\tilde{G}(1-f^2)}} e^h}\over {
(\prod_{p=1}^{p=n} |z-z_p|^2)^{\tilde{G}}
}}f^2 (2-f^2)\\
&\ge& c {{f^{2+2\tilde{G}}e^h }   \over { (\prod_{p=1}^{p=n}
|z-z_p|^2)^{\tilde{G}} }}\\
&\ge& c' {{f^{2+2\tilde{G}}e^h }   \over {  |z|^{2n\tilde{G}} }},
\eea \noindent where $c$ and $c'$ are some positive constants.
Since there is no solution of the  equation  $ \Delta w=K(z) e^w $
when $K(z)$ is positive and $K(z) \ge c' |z|^{-2}$ as $|z|\to
\infty$ (see \cite{Sat}),  we conclude that there is no solution
for $\psi$
 when $f^2 \ge c |z|^{ { {2(n\tilde{G}-1)}\over {1+
\tilde{G}}} }$ for any positive constant $c$ and large $|z|$.

\section*{Acknowledgments}
{The authors would like to thank Chanju Kim and Kyoungtae Kimm for
helpful discussions. This work was supported by KOSEF96070102013 and
KRF1998-015-D00034(S. Kim), No. 2000-1-11200-001-3 from the Basic Research
Program of the Korea Science $\&$ Engineering Foundation and
Korea Research Center for Theoretical Physics and Chemistry(Y. Kim).}


\begin{thebibliography}{100}
\bibitem{JT} R. Jackiw and S. Templeton, {\it Phys. Rev. D} {\bf 23} (1981),
2291; J. Schonfeld, {\it Nucl. Phys. B} {\bf 185} (1981), 157; S. Deser, R.
Jackiw, and S. Templeton, {\it Phys. Rev. Lett.} {\bf 48} (1982), 975;
{\it Ann. Phys. (N.Y.)} {\bf 140} (1982), 372.
\bibitem{ASWZ} D. Arovas, J. Schrieffer, F. Wilczek, and A. Zee, {\it Nucl.
Phys. B} {\bf 251} (1985), 117; C. R. Hagen, {\it Ann. Phys. (N.Y)} {\bf 157}, 342 (1984).
\bibitem{Wit} E. Witten, {\it Commun. Math. Phys.} {\bf 121} (1989), 351.
\bibitem{BN} M. Bos and V. P. Nair, {\it Phys. Lett. B} {\bf 233} (1989), 61.
\bibitem{WZ} F. Wilczek and A. Zee, {\it Phys. Rev. Lett.} {\bf 51} (1983),
2250; Y. S. Wu and A. Zee, {\it Phys. Lett. B} {\bf 147} (1984), 325.
\bibitem{PK} S. K. Paul, and A. Khare, {\it Phys. Lett. B} {\bf 174} (1986),
420.
\bibitem{HKP} J. Hong, Y. Kim, and P. Y. Pac, {\it Phys. Rev. Lett.} {\bf 64}
(1990), 2230; R. Jackiw and E. J. Weinberg, {\it ibid} {\bf 64} (1990), 2234.
\bibitem{JLW} R. Jackiw, K. Lee, and E. J. Weinberg, {\it Phys. Rev. D} {\bf 42}
(1990), 3488.
\bibitem{Dun} G. Dunne, {\it Self-dual Chern-Simons Theories} (Springer,
Berlin, 1995).
\bibitem{CG} B. Linet, {\it Phys. Lett. A} {\bf 124} (1987), 240; {\it Gen.
Rel. Grav.} {\bf 20} (1988), 451; A. Comtet and G. W. Gibbons, {\it Nucl.
Phys. B} {\bf 299} (1988), 719.
\bibitem{Sch} J. Schiff, {\it J. Math. Phys.} {\bf 32} (1991), 753;
A. Comtet and A. Khare, {\it Phys. Lett. B} {\bf 278} (1992), 236;
S. Kim and Y. Kim, Preprint math-ph/0012045.
\bibitem{Val} P. Valtancoli, {\it Int. J. Mod. Phys. A} {\bf 18} (1992), 4335;
D. Cangemi and C. Lee, {\it Phys. Rev. D} {\bf 46} (1992), 4768;
G. Clement, {\it Phys. Rev. D} {\bf 54} (1996) 1844.
\bibitem{ES} M. B. Einhorn and R. Savit, {\it Phys. Rev. D} {\bf 17} (1978),
2583; M. E.  Peskin, {\it Ann. Phys. (N.Y.)} {\bf 113} (1978), 122.
\bibitem{KL} Y. Kim and K. Lee, {\it Phys. Rev. D} {\bf 49} (1994), 2041.
\bibitem{Wan} R. Wang, {\it Commun. Math. Phys.} {\bf 137} (1991), 587;
J. Spruck and Y. Yang, {\it Commun. Math. Phys.} {\bf 149} (1992), 361.
\bibitem{DJ} S. Deser and R. Jackiw, {\it Phys. Lett. B} {\bf 139} (1994), 371;
S. Deser and Z. Yang, {\it Mod. Phys. Lett. A} {\bf 4} (1989), 2123.
\bibitem{KK} C. Kim and Y. Kim, {\it Phys. Rev. D} {\bf 50} (1994), 1040.
\bibitem{Lin} B. Linet, {\it Class. Quantum Grav.} {\bf 7} (1990), L75;
Y. Yang, {\it Commun. Math. Phys.} {\bf 170} (1995), 540.
\bibitem{DJH} A. Staruszkiewicz, {\it Acta. Phys. Polon.} {\bf 24} (1963), 735;
S. Deser, R. Jackiw, and G 't Hooft, {\it Ann. Phys. (N.Y.)} {\bf 152} (1984),
220; J. Gott and M. Alpert, {\it Gen. Rel. Grav.} {\bf 16} (1984), 243;
S. Giddings, J. Abbot, and K. Kuchar, {\it Gen. Rel. Grav.} {\bf 16} (1984),
751.
\bibitem{Hen} M. Henneaux, {\it Phys. Rev. D} {\bf 29} (1984), 2766.
\bibitem{LLW} C. Lee, K. Lee, and E. J. Weinberg, {\it Phys. Lett. B} {\bf 243}
(1990), 105.
\bibitem{BBS} E. Witten, {\it Int. J. Mod. Phys. A} {\bf 10} (1995), 1247;
K. Becker, M. Becker, and A. Strominger, {\it Phys. Rev. D} {\bf 51} (1995),
R6603; J. D. Edelstein, C. Nunez, F. A. Schaposnik, {\it Nucl. Phys. B}
{\bf 458} (1996) 165, {\it Phys. Lett. B} {\bf 375} (1996) 163.
\bibitem{KK2} Y. Kim and K. Kimm, {\it Phys. Rev. D} {\bf 58} (1998), 107701.
\bibitem{Gil} D. Gilbarg and N. Trudinger, {\it Elliptic Partial Differential
Equations of Second Order} (Springer, New York, 1983).
\bibitem{Sat} D. Sattinger, {\it Indiana Math. J.} {\bf 22} (1972), 1.
\end{thebibliography}
\end{document}